\documentclass{aa} 

\usepackage{graphicx}
\usepackage[section]{placeins}
\usepackage{scalerel}
\usepackage{siunitx}
\usepackage{hyperref}
\usepackage{makecell}
\usepackage{comment}
\usepackage{orcidlink}

\newcommand\HII{H\protect\scaleto{$II$}{1.2ex}}

\hypersetup{colorlinks=true,
            linkcolor=blue,
            citecolor=blue,
            urlcolor=blue}

\usepackage{txfonts}

\begin{document}

   \title{Filamentary accretion flows in high-mass star-forming clouds}

   \subtitle{}

   \author{J.-E. Schneider\inst{1,}\inst{2} \orcidlink{0009-0004-9578-0982}, H. Beuther\inst{1}, C. Gieser\inst{1}, S. Jiao\inst{1}, M.R.A. Wells\inst{1,}\inst{2}, 
         R. Klessen\inst{3}, S. Feng\inst{4} \orcidlink{0000-0002-4707-8409}, P. Klaassen\inst{5}, M. T. Beltran\inst{6}, R. Cesaroni\inst{6}, S. Leurini\inst{7}, J.S. Urquhart\inst{8}, A. Palau\inst{9} and R. Pudritz\inst{10}}

   \institute{Max-Planck Institut für Astronomie,
              K\"onigstuhl 17, D-69117 Heidelberg\\
              \email{jschneider@mpia.de}
        \and
            Fakultät für Physik und Astronomie, Universität Heidelberg, Im Neuenheimer Feld 226, D-69120 Heidelberg
        \and
            Zentrum für Astronomie der Universität Heidelberg, Albert-Ueberle-Str. 2, D-69120 Heidelberg 
        \and
            Department of Astronomy, Xiamen University (Haiyun Campus), Zengcuo'an West Road, Xiamen, 361995 China 
        \and
            UK Astronomy Technology Centre, Royal Observatory, Edinburgh, Blackford Hill, Edinburgh, EH9 3HJ, United Kingdom
        \and   
            INAF - Osservatorio Astrofisico di Arcetri, Largo Enrico Fermi 5, I-50125 Firenze, Italy
        \and
            INAF - Osservatorio Astronomico di Cagliari, Via della Scienza 5, I-09047 Selargius (CA), Italy
        \and
            Centre for Astrophysics and Planetary Science, University of Kent, Canterbury CT2 7NH, UK
        \and
            Instituto de Radioastronomía y Astrofísica, Universidad Nacional Autónoma de México, Antigua Carretera a Pátzcuaro 8701, Ex-Hda. San José de la Huerta, 58089 Morelia, Michoacán, México
        \and   
            Department of Physics and Astronomy, McMaster University, Hamilton, ON, Canada
             }

   \date{}

  \abstract
   {Filamentary accretion-flows as gas-funneling mechanisms are a key aspect in high-mass star formation research. The kinematic properties along these structures are of particular interest. }
   {The scope of this paper is focused on the question if gas is transported to dense clumps inside high-mass star-forming regions through filamentary structures on a scale of several parsecs to the sub-parsec scale. }
   {We quantify the gas flows from a scale of up to several parsecs down to the sub-parsec scale along filamentary structures. In this work the accretion flow mechanisms based on gas kinematic data in the three high-mass star-forming regions G75.78, IRAS21078+5211 and NGC7538 are studied with data obtained from the IRAM 30\,m telescope. The analysis is carried out using the surface density derived from 1.2\,mm continuum emission and velocity differences estimated from HCO$^+$\,($1-0$) and H$^{13}$CO$^+$\,($1-0$) molecular line data. }
   {The mass flow-behavior of the gas in the vicinity of high-mass star-forming clumps shows characteristic dynamical patterns, such as an increased mass flow rate towards the clumps. We assume the velocity differences to originate from filamentary-gas infall onto the high-mass star-forming clumps; the inclination of the filament-structures along the line of sight is however unknown. Nevertheless, using the velocity differences and mass surface densities, we can estimate mean flow rates along the filamentary structures with respect to the line of sight and towards clumps. We quantify the flow rates towards the clumps in a range from about $10^{-3}\,M_{\odot}\,$yr$^{-1}$ to $10^{-5}\,M_{\odot}$\,yr$^{-1}$, inferred from clump-centered polar plots. Slight variations of the flow rates along the filamentary structures may be caused by overdensities and velocity gradients along the filaments. }
   {While the initial studies presented here already reveal interesting results such as an increasing mass flow rate towards clumps, the properties of filamentary gas flows from large to small spatial scales, as well as potential variations over the evolutionary sequence, are subject to future studies.}


   \authorrunning{J.-E. Schneider et al.}
   \titlerunning{}
   \maketitle
%

\section{Introduction}

Understanding filamentary accretion-flows as a central mechanism in high-mass star formation is still an open field of research with many aspects yet unknown (e.g. \citealt{Hacar2023}, \citealt{Pineda2023}). Theory provides us with models of cloud-collapse, fragmentation and other processes that lead to filamentary accretion-flows in giant molecular clouds (GMC) and molecular clouds (MC) (e.g. \citealt{RSmith2012}, \citealt{RSmith2014}, \citealt{RSmith2016}, \citealt{Vazquez}, \citealt{Padoan}, \citealt{Zhao+2024}). The difference between GMCs and MCs is their mass: GMCs are larger and much more massive, exceeding $10^5\,M_{\odot}$. In this study, we focus on parsec-scale accretion-flows from the cloud environment onto the dense high-mass clumps prone to massive cluster formation.

\begin{figure*}
   \includegraphics[width=0.99\linewidth ,keepaspectratio] {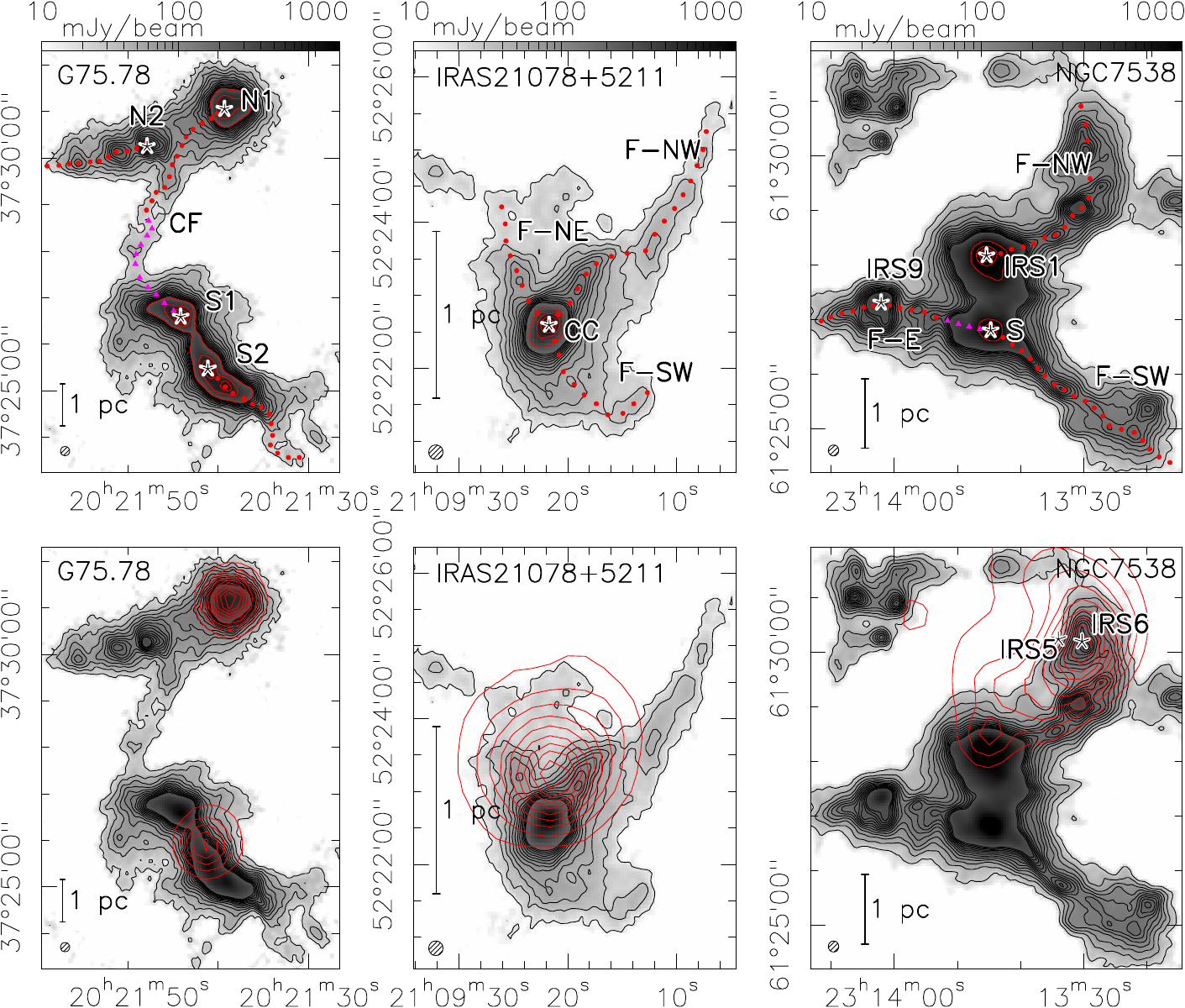}
      \caption[Naming and measuring pattern for the three sources]{Top row: NIKA2 1.2\,mm dust continuum maps (see \citealt{beuther2024}) of the star-forming regions showing the clump and filament designations. The red and magenta markers indicate the data points used for the flow rate estimation in Section \ref{sec:fila-flows}. Triangular magenta markers indicate a separate section along a structure. Black contours outline the 1.2\,mm continuum, ranging from 3$\sigma$ to 39$\sigma$ in steps of 3$\sigma$ (the RMS-values are presented in Table \ref{tab:dust_hii_rms}). The red contours in the upper panels mark the peak-positions of the continuum emission from 20\% to 100\% in steps of 20\%. Designations used in the following refer to clumps and filament-structures, filaments are marked by an ``F'' in the designation; ``CF'' in G75 stands for ``central filament'', ``CC'' in IRAS21078 refers to ``central clump'' while the clumps in NGC7538 are the known objects IRS9, IRS1 and S (e.g. \citealt{Beuther}). G75 S1 is the object G75.78+0.34. We present the clump-designations together with their coordinates in Table \ref{tab:clumps}. Bottom row: The same dust-maps with red contour lines outlining the \HII-regions at 1.4\,GHz (21\,cm) (\citealt{CondonVLA_1998}); the contours range from 10\% to 100\% of the peak emission in steps of 10\% (see Table \ref{tab:dust_hii_rms} for reference). The positions of the two main exciting sources of the \HII-region in NGC7538, NGC7538 IRS5 and NGC7538 IRS6, are marked in the lower right panel (\citealt{Puga}).}
        \label{fig:dusttags}
\end{figure*}

The three star-forming regions studied in this paper are  G75.78 with the main clump G75.78+0.34 (hereafter G75), IRAS21078+5211 (hereafter IRAS21078) and NGC7538, consisting of NGC7538 IRS1, NGC7538 S and NGC7538 IRS9. The sources are part of the CORE program (\citealt{Beuther2018}, \citealt{Beuther}, \citealt{Gieser2021}, \citealt{Ahmadi2023}, \citealt{beuther2024}). We summarize their main parameters in Table \ref{tab:source_params}. According to their estimated masses and equivalent radii of about 1\,pc, we refer to them as clumps throughout this work. They have been subject to studies in the past; young stellar objects (YSOs) and \HII-regions have been observed in all regions (see e.g. \citealt{Riffel}, \citealt{Moscadelli}, \citealt{Beuther2022}). While the initial aim was to include all CORE-regions into this survey, we only obtained data for three of them due to weather conditions during the observing window. All regions would have represented a broader range of mass regimes; this is why IRAS21078 is included in this work despite its lower mass compared to the other two sources.\par 

\begin{figure*}
    \centering
    \includegraphics[width=0.99\linewidth ,keepaspectratio] {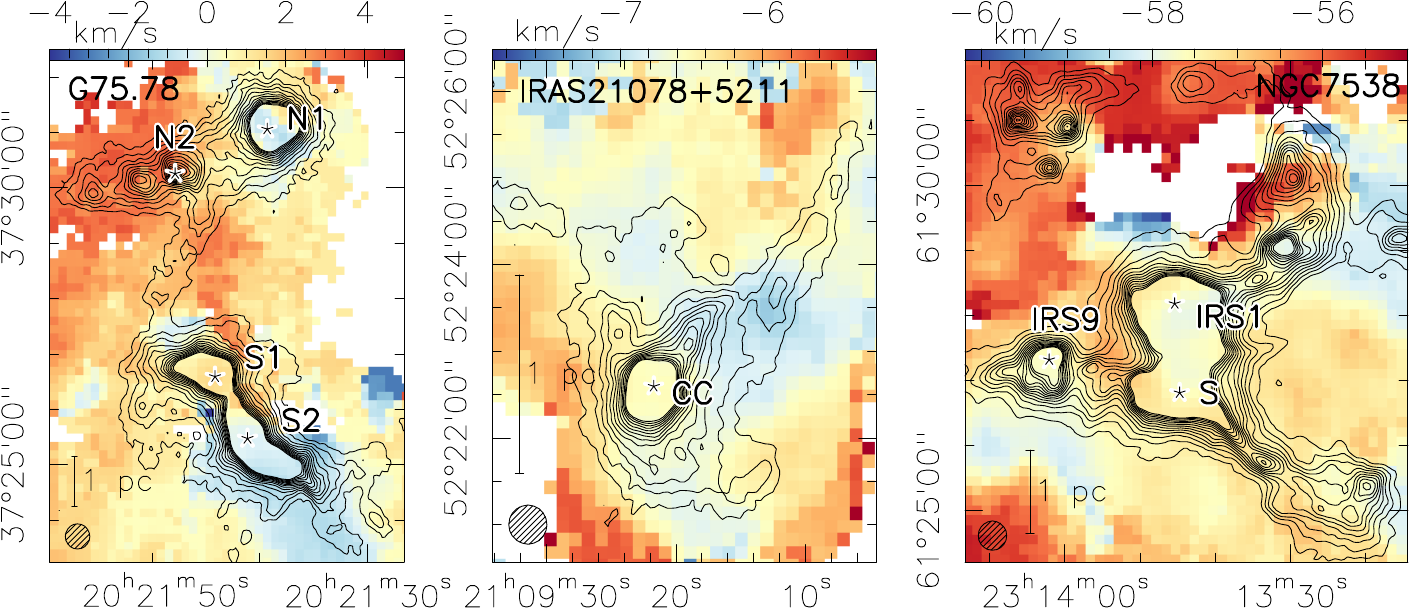}
    \caption[1st moment maps HCO$^+$\,($1-0$)]{First moment (intensity-weighted peak-velocity) maps for the HCO$^+$\,($1-0$)-line. The beam size is \SI{27}{\arcsecond} (lower left corner). The sources are labeled in each panel. The scale bar indicates the length of 1 parsec in each panel. Contour lines show the 1.2\,mm dust continuum ranging from 3$\sigma$ to 39$\sigma$ in 3$\sigma$-steps.}
    \label{fig:1stmoment_hco+}
\end{figure*}

\begin{figure*}
    \centering
    \includegraphics[width=0.99\linewidth ,keepaspectratio] {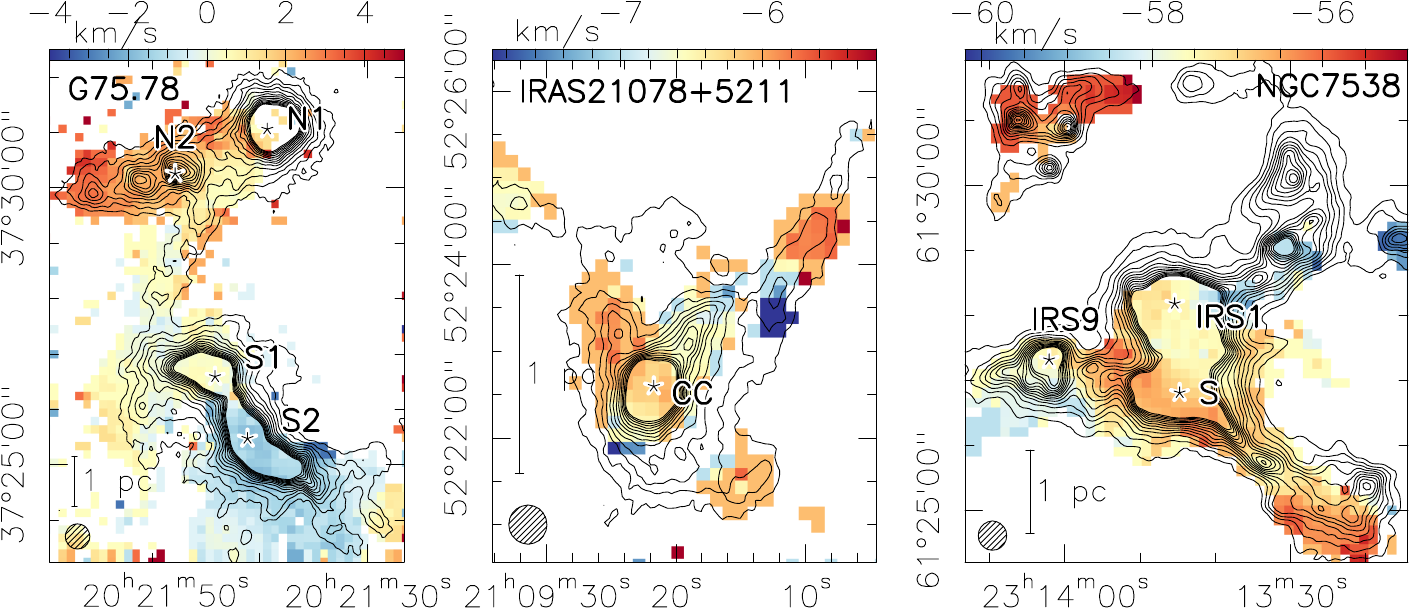}
    \caption[1st moment maps H13CO$^+$\,($1-0$)]{As for Fig. \ref{fig:1stmoment_hco+} but for H$^{13}$CO$^+$\,($1-0$)-transition.}
    \label{fig:1stmoment_h13co+}
\end{figure*}

Filament-structures and hubs have been studied in other star-forming regions in the past, and velocity gradients along filaments have been identified (e.g. \citealt{Peretto2013}, \citealt{Peretto2014}, \citealt{Kim2022}, \citealt{Yingxiu2023_filaments_accretion}). Whether observed velocity differences traced by molecules are signatures from flows inside filament-structures however is still not entirely clear. In this work we will investigate such structures in all three regions in detail and apply a simple mass flow estimation model. The mass flow approach to quantify accretion-flows in high-mass star-forming regions has already been outlined and used in previous studies (e.g. \citealt{Beuther}, \citealt{Wells2024}). On the theoretical side, processes from large to small scales have been, for example, modeled and categorized into inertial infall and subsequent inflow, followed by accretion onto prestellar clumps (e.g. \citealt{RSmith2012}, \citealt{RSmith2014}, \citealt{RSmith2016}, \citealt{Padoan}, \citealt{Zhao+2024}). On the other hand, the formation of filamentary structures inside GMCs and MCs is assumed to be a direct consequence of enhanced anisotropy during cloud collapse and fragmentation, after which the filaments act as channels were gas is funneled onto the prestellar clumps (e.g. \citealt{Vazquez}, \citealt{Padoan}). Most simulations show also that the formation of filaments is a consequence of crossing shocks (\citealt{Vazquez+2003}, \citealt{MacLowKlessen2004}, \citealt{Krumholz+2005}, \citealt{HennebelleChabrier2011}, \citealt{Padoan2011}, \citealt{FederrathKlessen2012}, \citealt{PudritzKevlahan2013}). The analysis of the high-mass star-forming MCs in this work generally takes into account scales of several parsecs down to the sub-parsec scale.\par
The structure of the paper is as follows: We will start by introducing the data in Sect. \ref{sec:obs}, also giving a brief overview of the reduction process and archival data. We then focus on the mass flow-analysis in Sect. \ref{sec:results}. We conclude this analysis with a summary and outlook in Section \ref{sec:conclusion_summary}.

\section{Observation}\label{sec:obs}

We present the NIKA2 1.2\,mm dust continuum emission maps for all three sources in Fig. \ref{fig:dusttags}. The first moment maps (intensity weighted peak-velocities) from HCO$^+$ and H$^{13}$CO$^+$ for the kinematic analysis are shown in Figs. \ref{fig:1stmoment_hco+} and \ref{fig:1stmoment_h13co+}, respectively. We summarize the RMS-values for the molecular data in Table \ref{tab:rms-molecules}. Data reduction and analysis have been conducted using the GILDAS\footnote{For further information, see \url{https://www.iram.fr/IRAMFR/GILDAS/}} framework with the subprograms CLASS and GREG.

\begin{table*}
\caption{Source parameters from the CORE sample (as listed in \citealt{Beuther2018}).}         
\label{tab:source_params}     
\centering                          
\begin{tabular}{l l l r r r r}       
\hline\hline                
    Source & RA (J2000) (h:min:s) & DEC (deg:$'$:$''$) & D (kpc) & M$^a$ (M$_\odot$) & $v_{\mathrm{lsr}}$ (km s$^{-1}$) & L ($10^4 \,$L$_\odot$)\\
     \hline 
     G75.78+0.34 & 20:21:44.03 & +37:26:37.70 & 3.8$^d$ & 13490 & -0.5$^d$ & 7.0$^f$ \\
     IRAS21078+5211 & 21:09:21.64 & +52:22:37.50 & 1.5$^b$ & 410 & -6.1$^b$ & 1.3$^b$ \\
     NGC7538 IRS1 & 23:13:45.36 & +61:28:10.55 & 2.7$^c$ & 12000 & -57.3$^{c,e}$ & 8.0$^f$ \\

     \hline 
\hline
\end{tabular}
\tablefoot{$^a$ The masses are estimated for the entire regions covered by the 1.2\,mm dust continuum maps assuming optically thin dust emission at a temperature of 20\,K (see also Section \ref{sec:estimate-math}).}
\tablebib{$^b$: \citealt{Molinari1996}, $^c$: \citealt{Reid+2009}, $^d$: \citealt{Ando+2011}, $^e$:\citealt{Moscadelli+2009}, $^f$:  RMS survey database (\url{http://rms.leeds.ac.uk/cgi-bin/public/RMS_DATABASE.cgi}).}
\end{table*}

\begin{table*}
\caption[Molecular species]{Overview of the molecular properties for the two molecules presented in this paper. }
\label{tab:mol-properties}    
\centering                    
\begin{tabular}{l l l l l l}   
\hline\hline             
   Molecule & Transition & Rest-freq. $\nu$ [GHz] & $A_{\text{ij}}$ [s$^{-1}$] & $E_{\text{u}}/k$  [K] & $n_{\text{crit}}$ [$\mathrm{cm}^{-3}]$\\
     \hline
     H$^{13}$CO$^+$ & 1-0 & 86.754 & $3.8\times10^{-5}$ & 4.2 & $4.1\times10^4$ \\
     HCO$^+$ & 1-0 & 89.189 & $4.2\times10^{-5}$ & 4.3 & $4.5\times10^4$ \\
\hline                                   
\end{tabular}
\tablefoot{Note that the critical density is for $T_k = 20$ K (kinetic temperature of colliding partners, \citealt{Shirley2015}). Einstein A coefficients and upper energy-levels are taken from the Splatalogue database (\url{https://splatalogue.online}). Both line data are are from JPL (Jet Propulsion Laboratory).}
\end{table*}

\begin{table}
\caption{Clump-designations and their coordinates.}           
\label{tab:clumps}    
\centering                         
\begin{tabular}{l c c}       
\hline\hline                
    Name & RA (J2000) (h:min:s) & DEC (deg:$'$:$''$) \\
    \hline
         G75 N1 & 20:21:39.3 & 37:31:03.6 \\
         G75 N2 & 20:21:47.9 & 37:30:15.5 \\
         G75 S1 & 20:21:44.2 & 37:26:37.5 \\
         G75 S2 & 20:21:41.2 & 37:25:28.6 \\
         IRAS21078 CC & 21:09:21.7 & 52:22:35.8 \\
         NGC7538 IRS1 & 23:13:45.4 & 61:28:10.8 \\
         NGC7538 S &  23:13:44.8 & 61:26:48.6 \\
         NGC7538 IRS9 & 23:14:02.1 & 61:27:19.0 \\
\hline                                   
\end{tabular}
\end{table}

\subsection{IRAM 30\,m EMIR observation}

We use three different sets of data in this analysis: The IRAM 30\,m spectral line data were observed in 2021 (project-id 121-20) as part of the CORE program (\citealt{Beuther2018}) with a beam size of \SI{27}{\arcsecond} in the 3\,mm band. The raw data was calibrated to the main-beam temperature $T_{\text{mb}}$ with a beam efficiency of 76\%. The spectral resolution is 0.9\,km s$^{-1}$. We identified the HCO$^+$-line at a frequency of $89.189$\,GHz in the upper sideband as the J=1-0 transition, with a $E_u/k$-value of $4.3$\,K and the H$^{13}$CO$^+$ emission as J=1-0 at $86.754$\,GHz with $E_u/k = 4.2$\,K. We list the molecular properties in Table \ref{tab:mol-properties}. The moment maps were produced with a lower intensity threshold of 4$\sigma$ (cf. Table \ref{tab:rms-molecules}). Because we noted some absorption features in the optically thick HCO$^+$ compared to the optically thin H$^{13}$CO$^+$ (Fig. \ref{fig:g75_spec}), we also included the latter molecular data into the analysis (Fig. \ref{fig:1stmoment_h13co+}).\par

We derived dust temperature maps by applying the iterative spectral energy distribution (SED) fitting procedure outlined by \citet{Lin2016ApJ}, using both Herschel data and the NIKA2 1.2\,mm maps. Prior to SED fitting, the NIKA2 1.2\,mm map was corrected for missing large-scale flux and free–free emission. Large-scale structures were recovered by combining the NIKA2 map with Herschel data in the Fourier domain using the J-comb algorithm (\citealt{Jiao2022SCPMA}). The combined map was then convolved to \SI{27}{\arcsecond} to match the angular resolution of the free–free emission map, which is derived based on H41$\alpha$ observations (for a detailed explanation, see \citealt{Gieser2023}, Section 4.2).
Finally, the free–free corrected NIKA2 1.2\,mm map was obtained by subtracting the free–free contribution from the combined map. Single-component modified blackbody fits were performed pixel by pixel after convolving all maps to a common beam size and regridding to the same pixel scale.
In the iterative fitting process (\citealt{Lin2016ApJ}), the final dust temperature maps were obtained at \SI{27}{\arcsecond} resolution.

\subsection{Archival continuum data}

We use 1.4\,GHz Very Large Array (VLA) data and temperature data from the Herschel infrared Galactic Plane Survey (Hi-GAL; \citealt{Molinari2016}, \citealt{Marsh}) and the IRAM 30\,m NIKA2 1.2\,mm dust continuum data (see e.g. \citealt{beuther2024}, Sect. 2.1). The underlying dust continuum maps that we use to identify filament-like structures originate from NIKA2 1.2\,mm continuum data from the CORE sample, observed with the IRAM 30\,m telescope (\citealt{beuther2024}) with a beam size of \SI{12}{\arcsecond}.\par
Radio continuum-data to probe the \HII-regions were taken from the NRAO VLA Sky Survey (NVSS) 1.4\,GHz VLA survey (see e.g. \citealt{CondonVLA_1998}), the peak and $\sigma_{\text{rms}}$-values of both the dust continuum-data and the \HII-data are presented in Table \ref{tab:dust_hii_rms}.

\section{Results}\label{sec:results}

In Fig. \ref{fig:1stmoment_hco+} we show the intensity-weighted peak velocities (1st moment) for HCO$^+$\,($1-0$) in the three sources, while we present the same for H$^{13}$CO$^+$\,($1-0$) in Fig. \ref{fig:1stmoment_h13co+}. The lower threshold when creating the maps was set to 3$\sigma$. We note interesting patterns: HCO$^+$\,($1-0$) and H$^{13}$CO$^+$\,($1-0$) display strong velocity differences of up to a few km\,s$^{-1}$ in the vicinity of N1 and prominently between S1 and S2 in G75. We also note gradients next to G75 CF and north of G75 S1. Inside IRAS21078, velocity differences of about 1 km\,s$^{-1}$ cover the extent of the cloud, with a sharp cutoff close to the eastern/south-eastern edge (except for H$^{13}$CO$^+$\,($1-0$), which is spatially confined to the denser regions of the source). In NGC7538, the velocity differences of $\sim$1 km\,s$^{-1}$ can be seen around IRS1 and S (cf. Fig. \ref{fig:dusttags}). What else besides gravitational infall could produce these velocity differences? Rotation of clumps appears an unlikely candidate since rotation of clouds on parsec scales has not convincingly been observed. The \HII-regions may impact the velocity structure, and one sees that in the bar-like structure in the north-west of NGC7538 (Fig. \ref{fig:1stmoment_hco+}). In G75, the southern \HII-region is located exactly in the middle of the clumps S1 and S2 (Fig. \ref{fig:dusttags}), it can hence even increase the gas flow towards the clump. Similar scenarios could happen in IRAS 21078. Hence, while the \HII-regions may have some influence, in the framework of this analysis, we assume that the velocity gradients are dominated by gas flows towards the clumps.\par

\begin{figure*}
    \centering
    \includegraphics[width=\hsize]{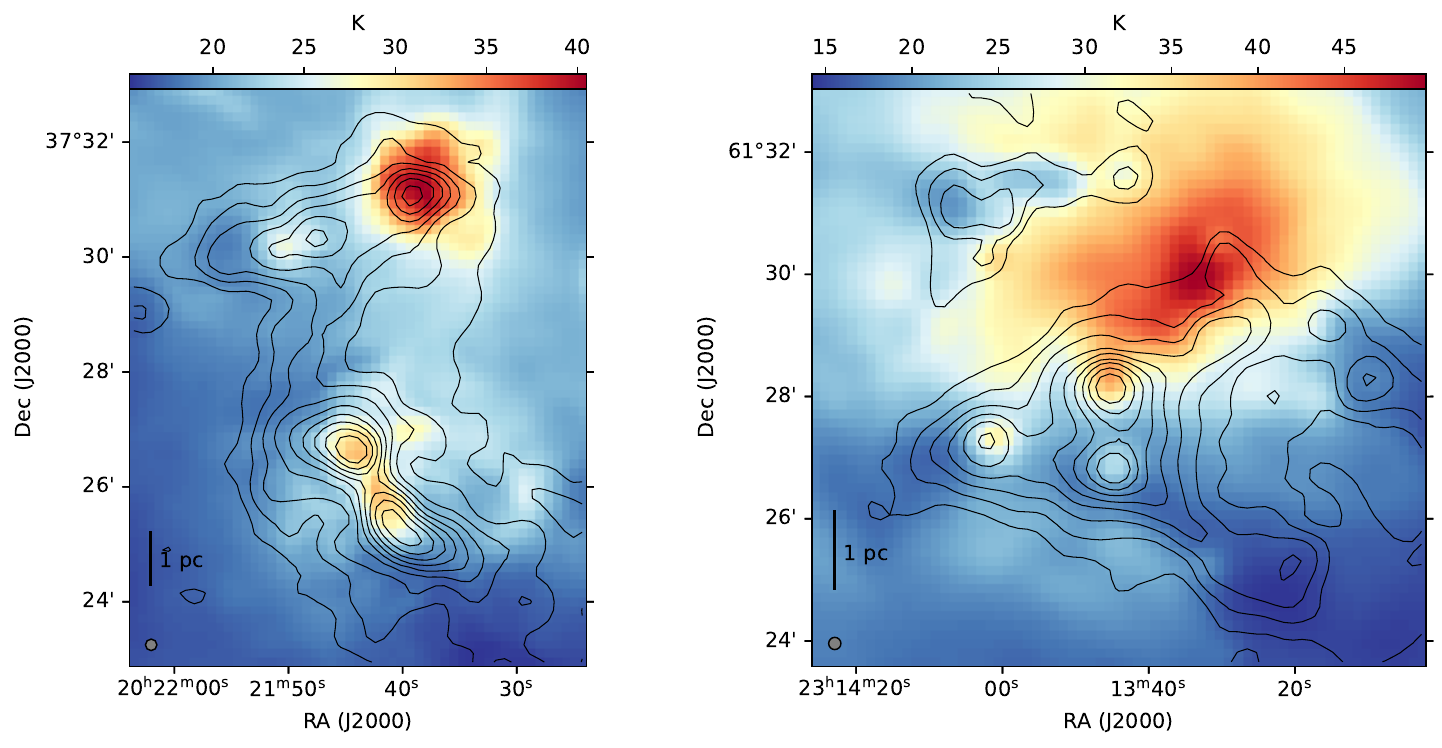}
    \caption[Temperature maps]{Dust temperature maps of G75 (left panel) and NGC7538 (right panel). Both maps were created using Herschel-data (\citealt{Molinari2010}) and free-free corrected dust continuum emission from the NIKA2 dataset. In the bottom left corner the \SI{27}{\arcsecond} beam of the final data product is shown. Black contours outline the free-free corrected 1.2\,mm continuum emission at \SI{27}{\arcsecond} resolution in logarithmic scale from 5$\sigma$ to peak value.
    \label{fig:g75_temp}}
\end{figure*}

We also note the lack of HCO$^+$-data in the northern part of NGC7538, but HCO$^+$\,($1-0$)-emission towards the peak of the \HII-region (cf. Figs. \ref{fig:dusttags}, \ref{fig:1stmoment_hco+} and \ref{fig:1stmoment_h13co+}). The \HII-region is excited by the two sources IRS5 and IRS6 with spectral types of O9 and O3 respectively (\citealt{Puga}) and already has photo-dissociated the molecular gas to a large extent (the other two sources have \HII-regions as well, but no lack of molecular emission from the HCO$^+$\,($1-0$)-line). The strong and sharp velocity differences of up to about 4 km\,s$^{-1}$ along the north-western filamentary structure in NGC7538 are therefore likely expected to stem from feedback from the \HII-region, and not from gas-infall. This assertion is important for the later interpretation of mass flow-behavior inside this source.\par
\begin{figure}
    \centering
    \includegraphics[width=\hsize]{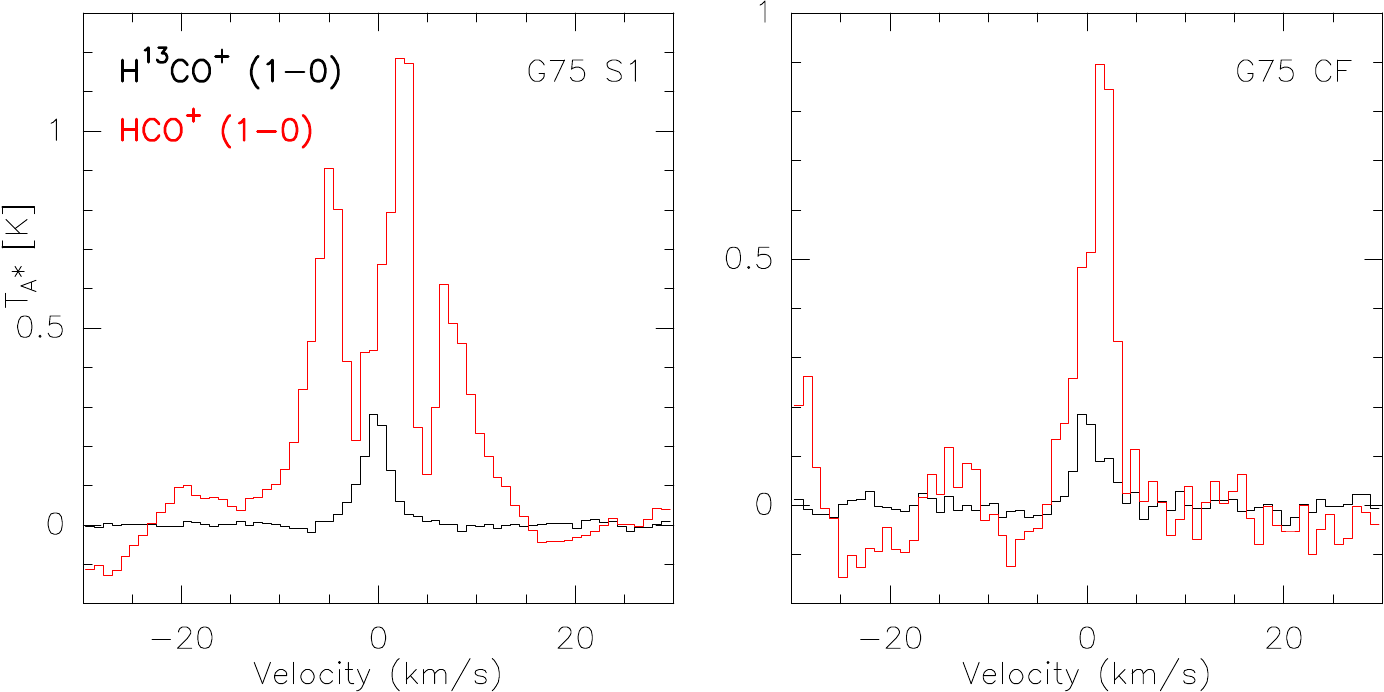}
    \caption[Spectra towards G75]{Spectra towards G75, showing molecular emission of HCO$^+$\,($1-0$) (red) and H$^{13}$CO$^+$\,($1-0$) (black). The left panel shows the emission at the clump position S1 in G75. The right panel shows the spectra for a position along the CF filament in G75 (see upper left panel of Fig. \ref{fig:dusttags}).}
    \label{fig:g75_spec}
\end{figure}

As shown in Figs. \ref{fig:g75_spec}, \ref{fig:spec_g75_s1} and \ref{fig:spec_g75_s2}, the main isotopologue can be optically thick towards the main emission peaks and may hence show multiple peaked spectra, whereas the rarer H$^{13}$CO$^+$ spectra always have Gaussian profiles, typical of optically thin lines. We do not estimate absolute optical depths because that is not needed for our analysis. Since we are interested in the peak velocities, we use the H$^{13}$CO$^+$ 1st moment maps to study the velocity field. Where H$^{13}$CO$^+$ is not detected, the optical depth of HCO$^+$ is probably low because it shows single Gaussian profiles. Therefore, at these positions the HCO$^+$ 1st moment maps can represent the velocity structure well.

\begin{table}
\caption{Overview of $\sigma_{\text{rms}}$-values of the molecular transitions at a spectral resolution of $0.9$\,km\,s$^{-1}$}            
\label{tab:rms-molecules}     
\centering                          
\begin{tabular}{l c}       
\hline\hline                
  Emission line & 1\,$\sigma_{\text{rms}}$ [mK]\\
     \hline
        HCO$^+$\,($1-0$) (G75) & 75 \\
        HCO$^+$\,($1-0$) (IRAS21078) & 56 \\
        HCO$^+$\,($1-0$) (NGC7538) & 78 \\
        H$^{13}$CO$^+$\,($1-0$) (G75) & 55 \\
        H$^{13}$CO$^+$\,($1-0$) (IRAS21078) & 43 \\
        H$^{13}$CO$^+$\,($1-0$) (NGC7538) & 60 \\

\hline                                 
\end{tabular}
\end{table}

\begin{table*}
\caption{Peak- and $\sigma_{\text{rms}}$-values of the 1.2\,mm dust continuum and the 1.4\,GHz (21\,cm) data (\HII-regions).}            
\label{tab:dust_hii_rms}      
\centering                         
\begin{tabular}{l c c c c }        
\hline\hline              
    Source & \thead{Peak 1.2\,mm \\ \scriptsize[Jy\,beam$^{-1}$]} & \thead{Peak 21\,cm \\ \scriptsize[Jy\,beam$^{-1}$]} & \thead{1\,$\sigma_{\text{rms}}$ 1.2\,mm \\ \scriptsize[mJy\,beam$^{-1}$]} & \thead{1\,$\sigma_{\text{rms}}$ 21\,cm \\ \scriptsize[mJy\,beam$^{-1}$]}\\
    \hline
         G75 & 1.45 & 5.00 & 7 & 5 \\
         IRAS21078 & 1.54 & 0.22 & 5 & 7.4 \\
         NGC7538 & 5.71 & 3.07 & 8 & 2.4 \\
\hline                           
\end{tabular}
\end{table*}

\subsection{Methods: flow rate estimates}\label{sec:estimate-math}

The estimation of the mass flow rate follows the approach used in \citealt{Beuther} and discussed in more detail by \citealt{Wells2024}. The filamentary structures are identified in the dust continuum emission and inclination $i$ relative to the observer is considered within the uncertainties discussed further below. We can estimate a mass flow rate as

\begin{equation}
    \dot{M}(i)=\Sigma_r\cdot \Delta v_r \cdot \frac{1}{\tan(i)} \cdot \Delta r_r, \label{eq:mdot}
\end{equation}

\noindent where $\Sigma_r$ is the beam-averaged mass surface density along the filamentary structures, $\Delta v_r$ is the velocity difference between a point and the center of measurement (a high-mass star-forming clump) and $\Delta r_r$ a projected distance along the structures and points of measurement. The mass flow rate estimate is a function of $i$, where $\tan(i)$ has to be non-vanishing; the subscript $r$ indicates the ``real'' values for the inclined filament-structure (for a detailed derivation, see \citealt{Wells2024}). The surface density can be estimated from the 1.2\,mm dust continuum assuming optically thin emission as (\citealt{Hildebrand}, \citealt{Schuller}):

\begin{equation}
    \Sigma= \frac{F_{\nu} R}{B_{\nu}(T_\text{D}) \kappa_{\nu}\Omega\mu m_\text{H}} \label{eq:sigma},
\end{equation}

\noindent where $F_{\nu}$ is the flux-density, $R$ the gas-to-dust ratio (150, \citealt{Draine2011}), $B_{\nu}$ the Planck function at the dust temperature $T_\text{D}$, $\kappa_{\nu}$ the dust absorption coefficient, which was set to 0.9\,$\mathrm{cm^2 \ g^{-1}}$ (as given in Table 1 col. 5 in \citealt{OssenkopfHenning}), $\mu$ the mean molecular weight of the interstellar medium; assumed to equal 2.8, $m_\text{H}$ the mass of the hydrogen atom and $\Omega$ the solid angle. We measure the flux-density $F_{\nu}$ from the 1.2\,mm dust continuum at 250\,GHz and consider the dust temperature obtained for G75 and NGC7538 from the Hi-GAL Galactic Plane Survey (\citealt{Molinari2016}) as mentioned in Sect. \ref{sec:obs}. As for IRAS21078, the temperature values were set constant to characteristic values of 20\,K, which has to be considered within the error margins. We checked the impact of the constant temperature with a comparison for G75 and NGC7538, where we averaged over the mass flow rates for each section of measurement and took the ratio of the variable temperature data and the constant temperature data. We estimated an average increase of the mass flow rate for the fixed values of about 75\%. We note that the difference in temperature close to peak-emission near the clumps is higher and therefore affects the overall mass flow rate to a larger degree. The systematic errors for the measured quantities in Eqs. (\ref{eq:mdot}) and (\ref{eq:sigma}) have been estimated. We assume a velocity uncertainty of 0.5 km\, s$^{-1}$, slightly larger than half the velocity resolution. From the SED fitting, each data point was weighted by its measured noise,
yielding an average dust temperature uncertainty of 1\,K. We furthermore estimated 5\% uncertainty for the flux-density (typical calibration uncertainty, see \citealt{Perotto2020}). Regarding the uncertainties of the projected distances to the center, since these are always relative distances within the maps, we assume an uncertainty of 1 pixel or \SI{5}{\arcsecond}. All errors have been estimated by applying Gaussian error propagation to Eqs. (\ref{eq:mdot}) and (\ref{eq:sigma}).

\begin{figure*}
    \centering
    \includegraphics[width=\hsize]{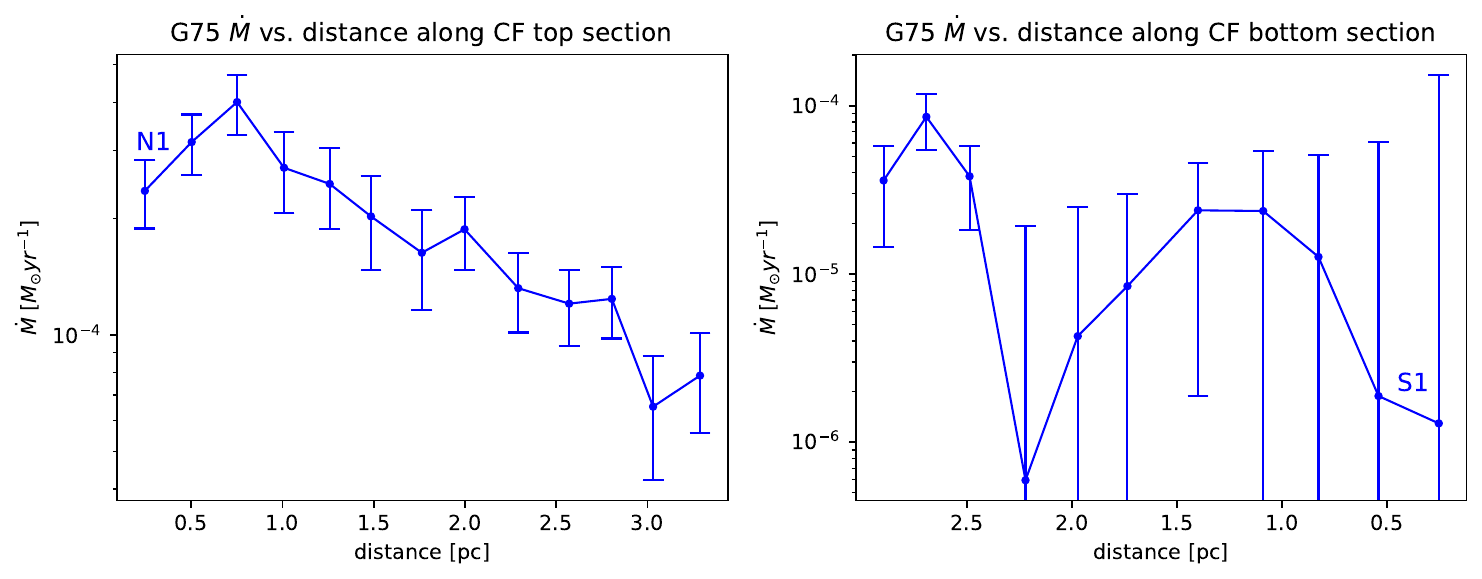}
    \caption[mass flow along G75 CF]{Estimated mass flow rate as a function of the distance along the CF filament structure in G75 (the connecting filament, see red and purple markers in Fig. \ref{fig:dusttags}). Left panel: top section. Right panel: bottom section. The plots follows the north-south direction from left to right. The velocity differences are taken relative to N1 for the left panel and relative to S1 for the right panel (see annotations).}
    \label{fig:g75_mf_cf}
\end{figure*}

\subsection{Flows along filaments}\label{sec:fila-flows}

\begin{figure}
    \centering
    \includegraphics[width=\hsize]{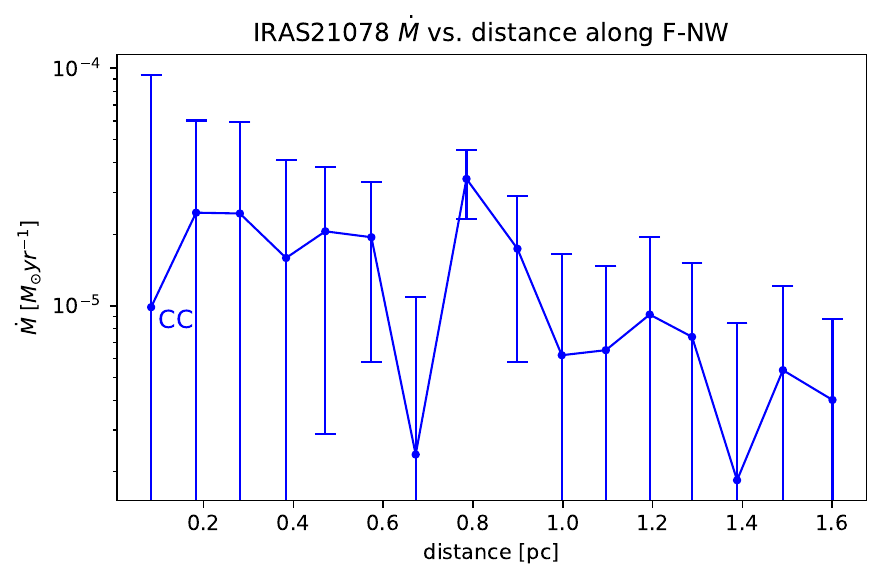}
    \caption[mass flow rate along IRAS21078 F-NW]{Estimated mass flow rate as a function of the distance along the F-NW-filament structure in IRAS21078 (Fig. \ref{fig:dusttags}). The graph follows the direction of measurement away from the clump, as indicated by the annotation.}
    \label{fig:iras_nwf}
\end{figure}

\begin{figure}
    \centering
    \includegraphics[width=\hsize]{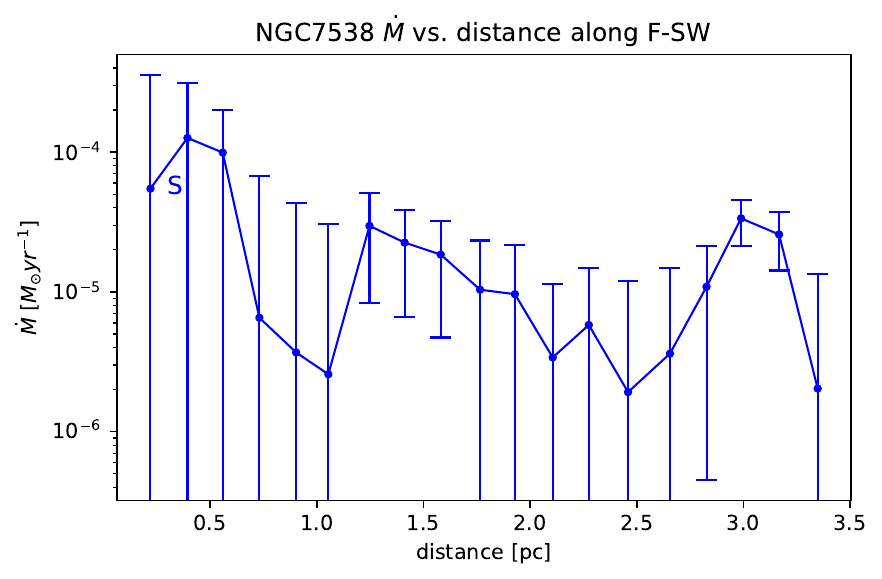}
    \caption[mass flow rate along NGC7538 F-SW]{mass flow rate as a function of the distance along the F-SW-filament structure in NGC7538 (Fig. \ref{fig:dusttags}). The graph follows the direction of measurement towards the south-west away from S, as indicated by the annotation.}
    \label{fig:ngc_fil_f-sw}
\end{figure}

For this part of the analysis we conducted the flux-density and velocity difference measurements along filamentary structures identified in the dust continuum maps. We plotted the mass flow rate as a function of the distance along the structure (see Figs. \ref{fig:g75_mf_cf}, \ref{fig:iras_nwf}, \ref{fig:ngc_fil_f-sw}). The mass flow rate was estimated with the separation $\Delta r$ as the half beam size of \SI{13.5}{\arcsecond} from the kinematic data and the surface density calculated from the flux-value taken at one marker-position, whereas the velocity difference was always taken relative to the center of the starting point (a clump, see also Section \ref{sec:estimate-math}). All velocity-values were taken preferably from H$^{13}$CO$^+$ emission, and only from HCO$^+$ emission if the former was undetected at the respective position in the map (Fig. \ref{fig:dusttags}).\par

\begin{figure*}
\centering
\includegraphics[width=\hsize]{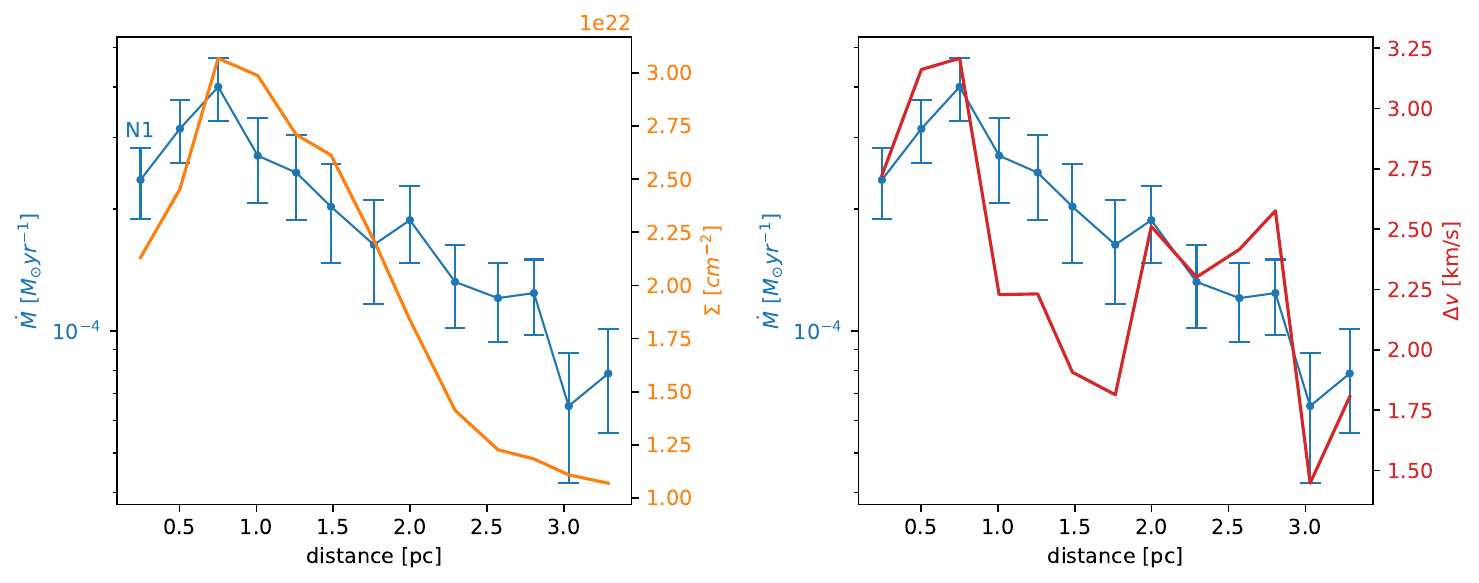}
 \caption[Comparison of contributors to $\dot{M}$ in G75 CF top section]{Light blue graph: mass flow rate as a function of the distance between adjacent points in G75 CF top section (Fig. \ref{fig:dusttags}, left panel in Fig. \ref{fig:g75_mf_cf}). Orange graph in the left panel: surface density along the CF-structure top section in G75. Red graph in the right panel: velocity differences along the CF-structure top section in G75.}
 \label{Fig:g75_fil_cf_comp_velograd_sigma_1}
\end{figure*}

\begin{figure*}
\centering
\includegraphics[width=\hsize]{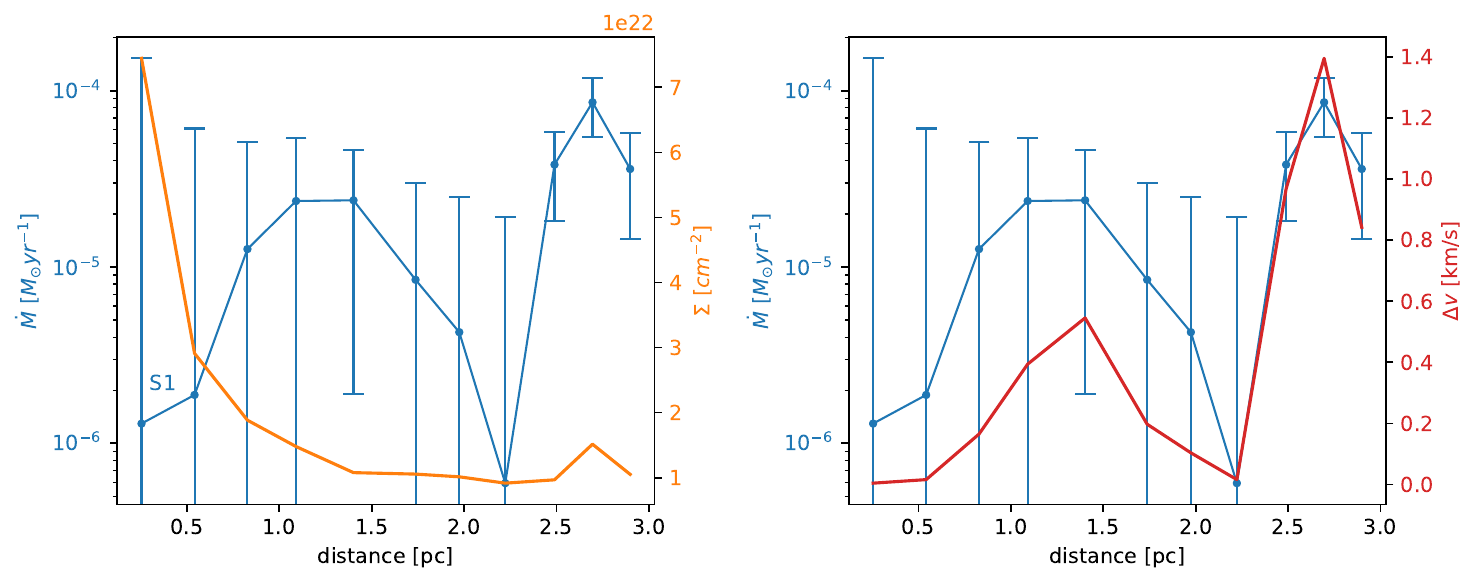}
 \caption[Comparison of contributors to $\dot{M}$ in G75 CF bottom section]{As for Fig. \ref{Fig:g75_fil_cf_comp_velograd_sigma_1} but for the CF bottom section.}
 \label{Fig:g75_fil_cf_comp_velograd_sigma_2}
\end{figure*}

We chose the main connecting filament in G75 (CF) and the northern and southern halves of the cloud. In IRAS21078, we focused on the three structures that stretch away from the central clump (CC) in north-eastern, north-western and south-western direction respectively (Fig. \ref{fig:dusttags}). In NGC7538, we picked the northern strip of the cloud, the eastern and the south-western structures (Fig. \ref{fig:dusttags}). As outlined in the previous section, the data do not directly yield information about the flow-direction. Following the picture of dynamical, filamentary clouds that collapse and feed star-forming gas clumps (e.g. \citealt{Myers2009}, \citealt{RSmith2014}, \citealt{RSmith2016}, \citealt{Vazquez}, \citealt{Padoan}) we assume that the flow is always directed towards the clumps. While outflows or expanding \HII-regions might in some region invert the flow direction, for our analysis we assume always that the flows are dominated by accretion flows towards the clumps. Because of this assumption, measurements along structures connecting two clumps were split, where the velocity differences where taken relative to one clump for one half, and relative to the other clump for the other half. This was conducted for CF in G75 and along F-E between IRS9 and S in NGC7538. In order to emphasize this separation, the relevant parts are marked by triangular magenta markers, as seen in Fig. \ref{fig:dusttags}. We note that all marker-positions were identified by eye. The same holds for the splitting of filaments.\par

In Fig. \ref{fig:g75_mf_cf} we show the mass flow rates along the two halves of the central filament (CF) structure in G75 (Fig. \ref{fig:dusttags}). We can see a characteristic shape of the curve along the structure. While in general the estimated flow rates in this filament decrease with increasing distance to the clump, we note a ``bump'' at approximate 2.25 to 2.75\,pc in the plot for the lower part. The same holds for the mass flow rate plot eastwards of N2 in G75 (Fig. \ref{fig:g75_mf_north}), where we also observe similar maxima. As for the southern part, the flow rate values from S2 southwards show a comparable irregular behavior with increasing distance. As for N2 and S2, we note nearby density-enhancements which stand out in the 1.2\,mm dust continuum (Fig. \ref{fig:dusttags}) and cause a large increase in surface density.\par

In Fig. \ref{fig:iras_nwf}, we present the estimated mass flow rate along the largest filament structure in IRAS21078, which shows an overall decline in magnitude with increasing distance from the central clump. The mass flow rate does, however, show intriguing peaks at about 0.7, 0.8 and 1.4\,pc respectively. We also can see strong depressions at about 0.6\,pc and 1.4\,pc in Fig. \ref{fig:iras_nwf}.\par

In Fig. \ref{fig:ngc_fil_f-sw} we present the same for the F-SW filament in NGC7538 (cf. Fig. \ref{fig:dusttags}), with again declining and peaking mass flow rate estimates along the structure with prominent depressions at about 0.5 to 1.2\,pc. All mentioned flow rate increases fall onto the position of overdense sections or minor clumps along filament-structures (in comparison of the flow rate graphs with Fig. \ref{fig:dusttags}), but at the same time we also can observe velocity gradients in the vicinity of these clumps (see Figs. \ref{fig:1stmoment_hco+} and \ref{fig:1stmoment_h13co+}). The trend of an increasing mass flow rate towards the denser clump-sections may therefore be partially associated with an increase in surface density. These sharp maxima along the mass flow rate curve are likely the result of steep velocity differences along the filament.\par

The mass flow rate estimation along filamentary structures was carried out for all three structures in IRAS21078 (Figs. \ref{fig:iras_nwf} and \ref{fig:iras_fil_ne-f}) and for the north-western, eastern and south-western structures in NGC7538 (Figs. \ref{fig:ngc_fil_f-sw}, \ref{fig:ngc_fil_f-e}, \ref{fig:ngc_fil_f-nw}). Most of the mass flow plots along filamentary structures show no clear decline in mass flow rate with increasing distance.\par

To separate important factors contributing to the mass flow rate, the mass surface density $\Sigma$ and the velocity difference $\Delta v$ have been plotted versus the distance in Figs. \ref{Fig:g75_fil_cf_comp_velograd_sigma_1} and \ref{Fig:g75_fil_cf_comp_velograd_sigma_2} to give a clearer distinction and indicate the more dominant contributor to $\dot{M}$. We can see that the overall shape of the curve is dominated by the surface density for the northern half of the filament and interestingly by the velocity differences for the southern half. Sharp features (i.e. maxima) along the curves stem from strong velocity differences. The possible cause for these velocity differences and the different impact of the two factors in the top and bottom part of G75 CF will be assessed in Sect. \ref{sec:conclusion_summary}.\par 

To be able to compare the correlation of velocity difference and surface density and their impact onto the mass flow rate generally, we plotted the separated factors of Eq. (\ref{eq:mdot}) for all major filament-structures that we analyzed in this sections and present them in Appendix \ref{sec:ap-a}. Many of the plots show spikes along the curve that either are caused by locally confined overdensities along filaments or steep variations of velocity differences. Most of the curves show a decline in surface density with increasing distance and spikes being caused by $\Delta v$ (see e.g. Figs. \ref{Fig:iras21078_fne_comp_mfcoldens}, \ref{Fig:iras21078_fnw_comp_mfcoldens}, \ref{Fig:iras21078_fsw_comp_mfcoldens}, \ref{Fig:ngc7538_fsw_comp_mfcoldens}, \ref{Fig:ngc7538_fnw_comp_mfcoldens}). In the case of the structure eastwards from N2 in G75, the surface density shows two maxima for two local overdensities (Fig. \ref{Fig:g75_n2_comp_mfcoldens} and Fig. \ref{fig:dusttags} for reference) but the overall mass flow rate only exhibits a maximum at about 1.7\,pc and not at about 0.7\,pc, where the velocity difference is low (between 0.2 and 0.3\,km\,s$^{-1}$ at 0.7\,pc compared to about 1\,km\,s$^{-1}$ at 1.7\,pc).\par
As for NGC7538 F-E, the mass flow rate is described by the velocity difference to a large degree (Fig. \ref{Fig:ngc7538_few_comp_mfcoldens}), whereas a sharp drop in surface density between 0.1 and 0.3\,pc does not visibly affect the course of the curve. In the case of the F-NW structure in NGC7538 (Fig. \ref{Fig:ngc7538_fnw_comp_mfcoldens}) the \HII-region affects the velocity-distribution of the molecular gas (e.g. \citealt{Beuther2022}) and thus causing the spikes in the velocity differences, excluding this graph for the accretion-analysis towards IRS1. The same holds for G75 S2 (\ref{Fig:g75_s2_comp_mfcoldens}), where the \HII-region is located nearby between S2 and S1 (see Fig. \ref{fig:dusttags} lower panels). A similar feature appears in the graph for the velocity difference for IRAS21078 F-NE (Fig. \ref{Fig:iras21078_fne_comp_mfcoldens}), where the sharp maximum at 0.4\,pc is located close to the center of the \HII-region (see Fig. \ref{fig:dusttags} bottom panels).

\subsection{Flows onto clumps}\label{sec:polar-flows}

In order to characterize the gas flow onto the clumps from all directions, the mass flow rates estimated with Eq. (\ref{eq:mdot}) are depicted in the form of polar plots for the central massive clumps and star-forming regions to give an intuitive picture of the two-dimensional spatial distribution of mass flow magnitudes. In every polar diagram we included a red dot as an indicator for the adjacent filament structure. These dots have been set by eye. The velocity differences have been evaluated by taking the value from one position around the center and subtract the given velocity-value for the center of the clump, similar to our approach along filaments (Section \ref{sec:fila-flows}). The distance $\Delta r$ in Eq. (\ref{eq:mdot}) was set to the half beam size of \SI{13.5}{\arcsecond} and the flux-values were taken for the same data point as the velocity, in order to be consistent with the previous flow rate-approach along the filaments.\par

\begin{figure}
    \centering
    \includegraphics[width=\hsize]{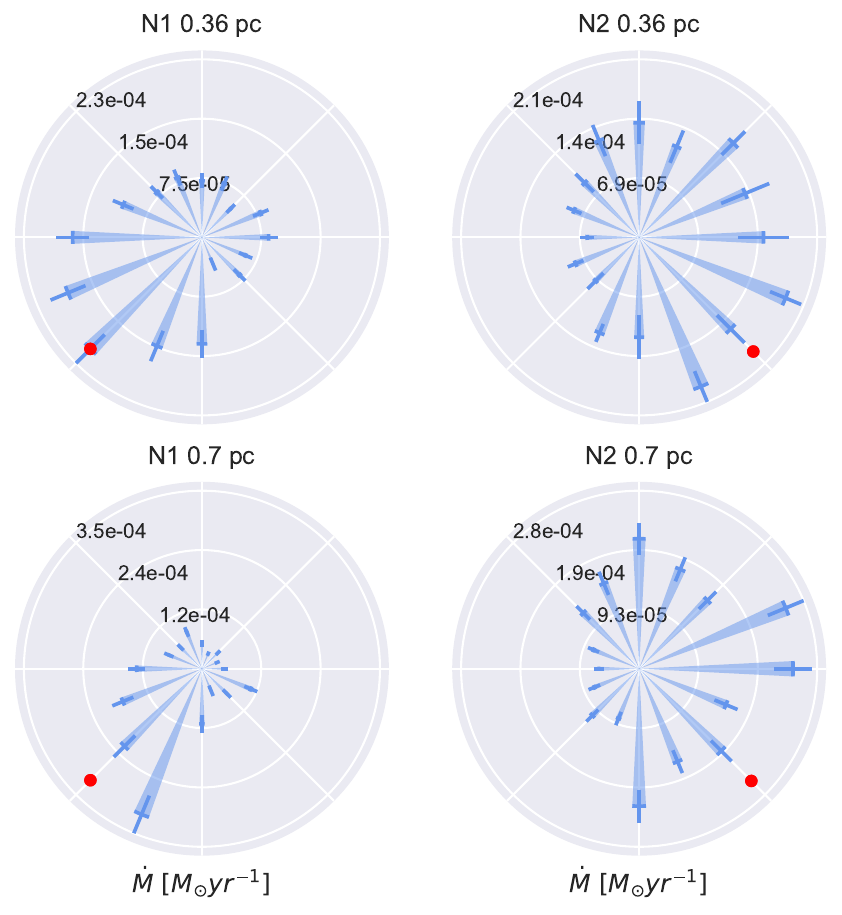}
    \caption[G75 northern clumps polar diagram]{Polar plots showing estimated mass flow rates as a function of the polar angle for the northern clumps in G75 (Fig. \ref{fig:dusttags}). Left column: mass flow rates towards N1, right column: mass flow rates towards N2. Top panels for the inner circle at 0.36\,pc, bottom panels for the outer circle at 0.7\,pc. Red dots indicate the direction of the CF-filament.}
    \label{fig:g75_mf_pol_n}
\end{figure}

In this way, the plots show mass flow rate-estimates always relative to the center, were 16 individual measurements were conducted (one data point every 22.5 degrees). The radial distances for the measurements were chosen to 20 and 40 arcseconds, resulting in a maximum distance to one clump of about 0.7\,pc for G75, 0.52\,pc for NGC7538 and 0.29\,pc for IRAS21078. Distances further out exhibit no significant surface density from the dust and are therefore not usable for this polar mass flow rate estimation (cf. Fig. \ref{fig:dusttags}). \par 

\begin{figure}
    \centering
    \includegraphics[width=\hsize]{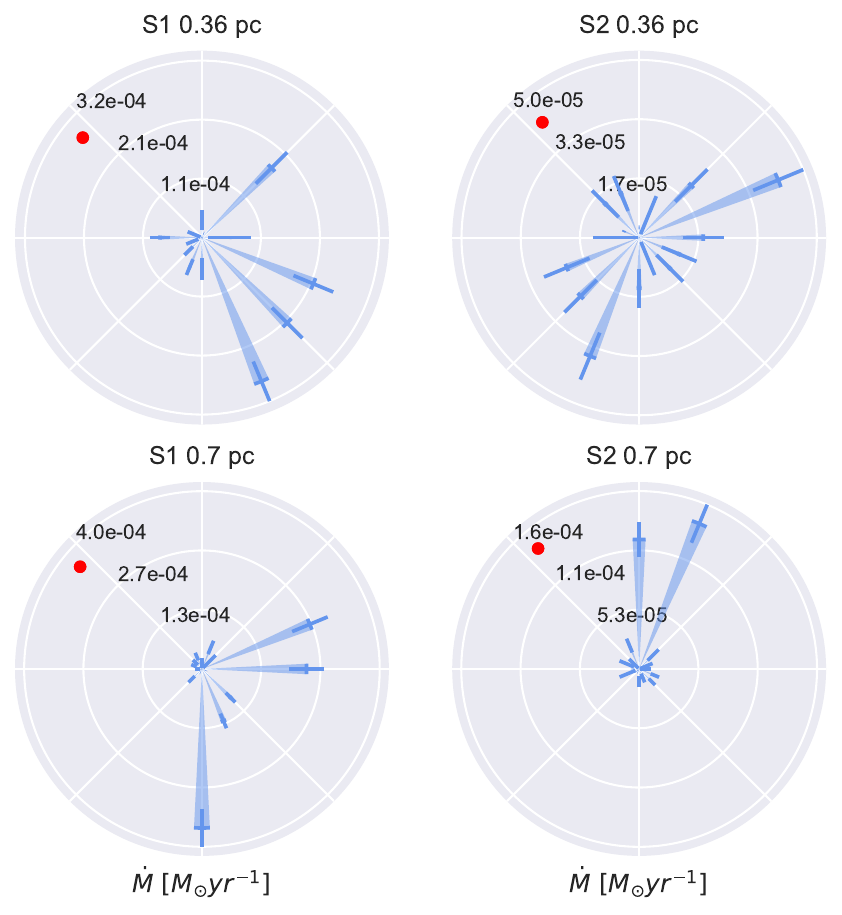}
    \caption[G75 southern clumps polar diagram]{As for Fig. \ref{fig:g75_mf_pol_n}, but for the southern clumps.}
    \label{fig:g75_mf_pol_s}
\end{figure}

For this study we investigated prominent clumps of star formation hosted inside the three clouds. All clumps have been identified by eye from the dust continuum maps. Their coordinates are given in Table \ref{tab:clumps}. These coordinates have been taken from the marker positions towards the clumps (see Fig. \ref{fig:dusttags}), which correspond to the peak positions of the dust continuum. for G75, we chose two regions in the northern part of the cloud (N1 and N2, see Fig. \ref{fig:dusttags}) and two in the southern part (S1 and S2, see Fig. \ref{fig:dusttags}). The two southern clumps are of particular interest, since there is a \HII-region between them (see. Fig. \ref{fig:dusttags}). In IRAS21078, we evaluated the mass flow-behavior at the central part of the cloud (CC, see Fig. \ref{fig:dusttags}) and in NGC7538 for IRS1, S and IRS9 (see Fig. \ref{fig:dusttags}). The flow rate estimation around prominent clumps provides further insight into the nature of mass accretion on sub-parsec scales. 

\begin{figure}
    \centering
    \includegraphics[width=\hsize]{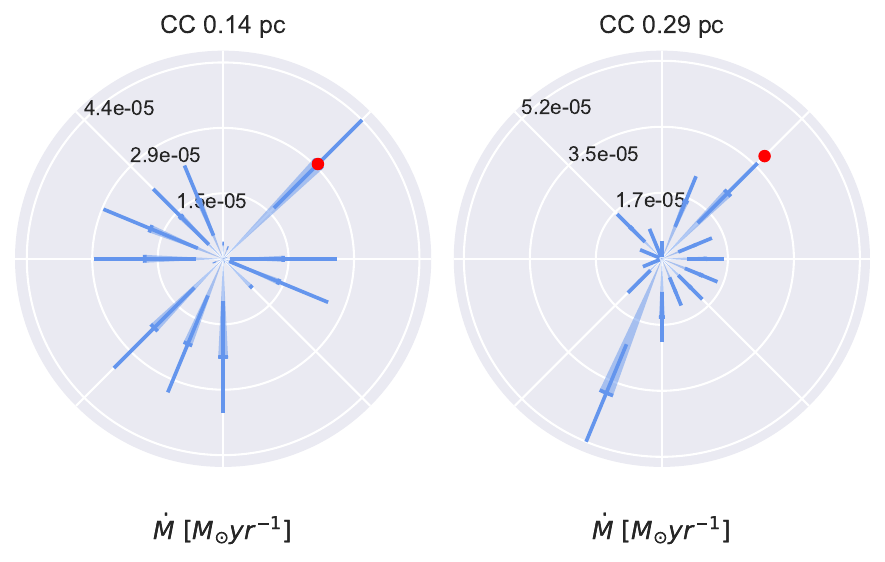}
    \caption[IRAS21078 CC polar diagram]{Polar mass flow diagram for the central clump (CC) in IRAS21078 (Fig. \ref{fig:dusttags}). Left panel for the inner circle at 0.14\,pc; right panel for the outer circle at 0.29\,pc. The red dot indicates the direction of the F-NW-filament.}
    \label{fig:iras_mf_pol}
\end{figure}

In Fig. \ref{fig:g75_mf_pol_n} we show the flow magnitudes in G75 between 20 and 40 arcseconds spacing for N1 and N2 respectively (cf. Fig. \ref{fig:dusttags}). The N1-panel shows only little orientation towards the filament for both distances, but seems to be more influenced by the eastern direction, towards N2 (Fig. \ref{fig:g75_mf_pol_n}). The N2-plot depicts a dominating mass flow rate in N1-direction for close proximity, and an increasing contribution from the east for increasing distance. The southern clumps are shown in Fig. \ref{fig:g75_mf_pol_s} without a clear orientation towards CF. S1 interestingly shows a strong orientation in mass flow rate magnitude towards S2, and none in direction of CF. S2 on the other hand seems to be primarily influenced from the direction of neighboring S1 and to the west for close proximity. We assume that both S1 and S2 are influenced by the \HII-region in between them (Fig. \ref{fig:dusttags}).\par

\begin{figure}
    \centering
    \includegraphics[width=\hsize]{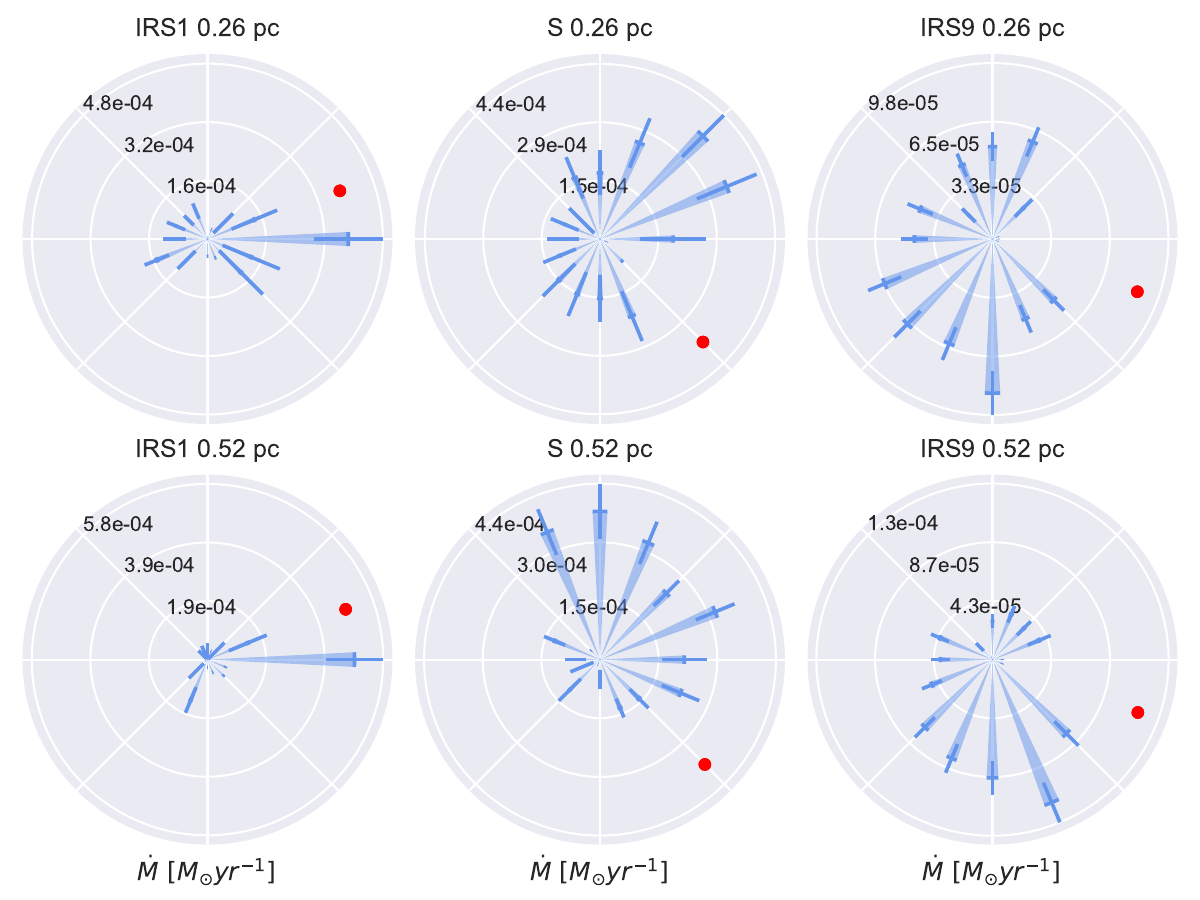}
    \caption[NGC7538 IRS1 and -S polar diagram]{Polar mass flow diagrams for NGC7538 IRS1, S and IRS9 (Fig. \ref{fig:dusttags}). Left panels IRS 1, middle panels S and right panels IRS9. Top panels for the inner circle of 0.26\,pc; bottom panels for the outer circle at 0.52\,pc. Red dots indicate the direction of the F-NW-filament (top row for IRS1) and the F-SW-filament (bottom row for S).}
    \label{fig:ngc7538_mf_pol}
\end{figure}

\begin{figure}
\centering
\includegraphics[width=\hsize]{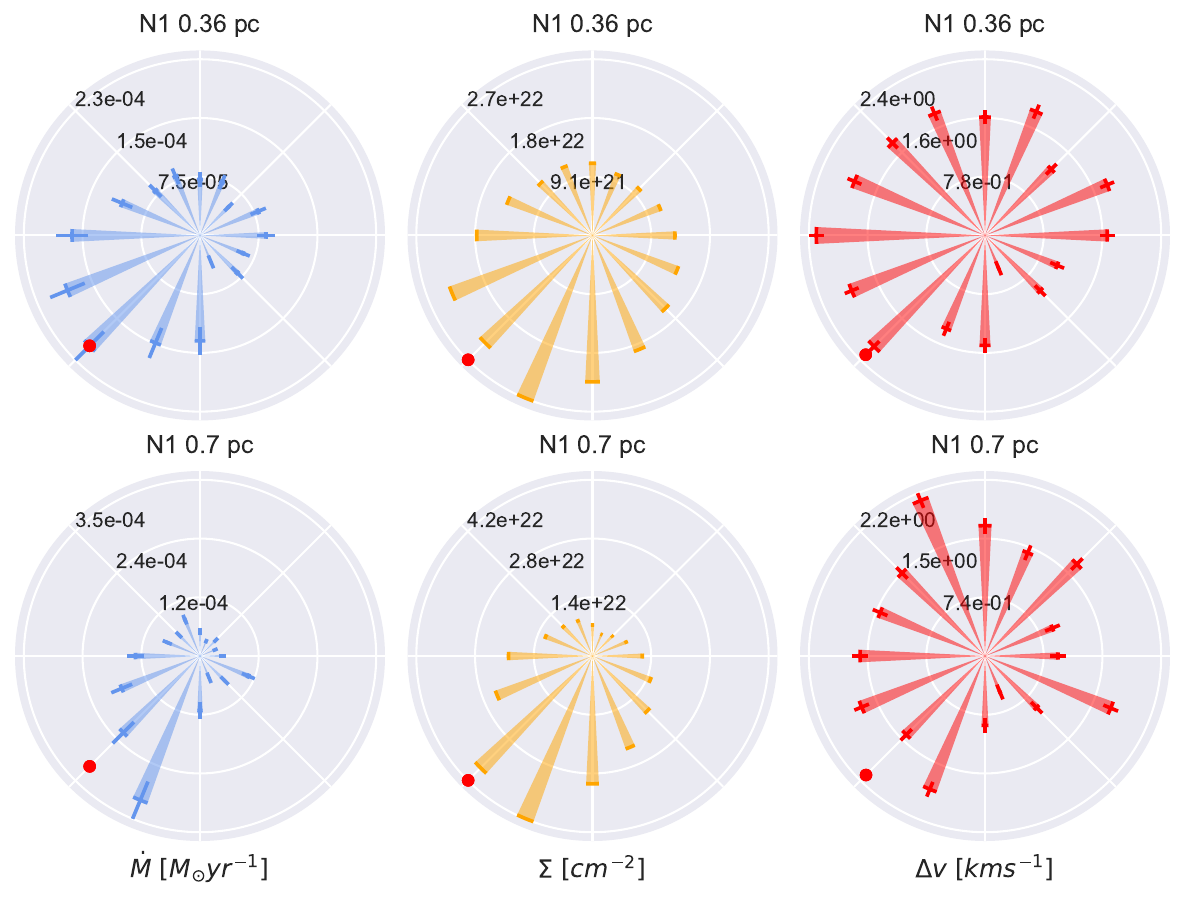}
 \caption{Comparison panels for the clump-centered flow approach. Columns from left to right: Total estimated mass flow rate, surface density and velocity difference for G75 N1 (Fig. \ref{fig:dusttags}). Top row: Inner circle at 0.36\,pc. Bottom row: Outer circle at 0.7\,pc. The red dots indicate adjacent filaments that are assumed to host the potential flow-direction.}
 \label{fig:g75_pcomp_vsigma_n-1}
\end{figure}

The diagram for IRAS21078 presents itself relatively featureless for 0.14\,pc distance (Fig. \ref{fig:iras_mf_pol}), but one can identify an increased magnitude of the mass flow rate to the north-west - in the direction of the largest filamentary structure. For increasing distance, an influence from the south-eastern edge becomes dominant however. The cloud has already been analyzed within the CORE-sample and the analysis has shown that it hosts converging flows of material towards its center (\citealt{Moscadelli}). The here presented mass flow direction does imply a potential gas-infall, but the channeling through the filaments has to be looked at in more detail. Comparing the two-dimensional spatial direction from the mass flow magnitudes in Fig. \ref{fig:iras_mf_pol} to the first moment maps in Figs. \ref{fig:1stmoment_hco+} and \ref{fig:1stmoment_h13co+} shows velocity differences along the structure of HCO$^+$ and H$^{13}$CO$^+$ prominently in the north-western direction, towards the elongated filament-structure. We also note strong HCO$^+$ velocity differences at the southern edge of the cloud (none for H$^{13}$CO$^+$ because of missing data). This possibly indicates infalling material, which is transported laterally onto the structure. The seemingly increasing velocity gradient towards blue-shifted gas at more densely packed contour lines of the dust continuum may indicate increasing dynamics due to channeling of the material closer to the center. This is however speculative as this assessment largely rests upon the HCO$^+$ and H$^{13}$CO$^+$ first moment maps alone. One always has to consider the presence of the \HII-regions, which likely cause alterations in potential gravitationally-driven infall via feedback-processes. This has to be noted especially for IRAS21078, where the extent of the \HII-region covers almost the whole source (Fig. \ref{fig:dusttags} lower panel). \par

\begin{figure}
\centering
\includegraphics[width=\hsize]{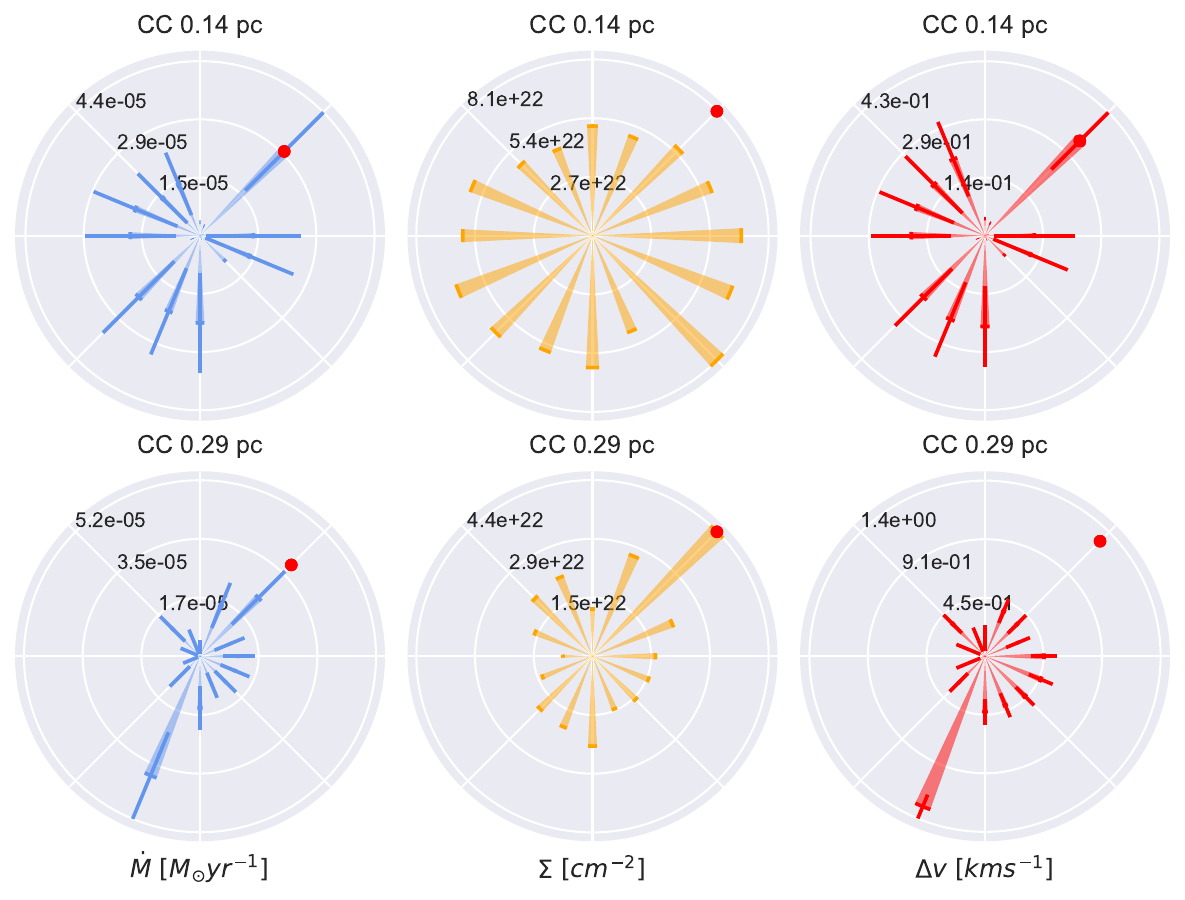}
 \caption{Comparison panels for the clump-centered flow approach. Columns from left to right: Total estimated mass flow rate, surface density and velocity difference for IRAS21078 CC (Fig. \ref{fig:dusttags}). Top row: Inner circle at 0.14\,pc. Bottom row: Outer circle at 0.29\,pc. The red dots indicate adjacent filaments that are assumed to host the potential flow-direction.}
 \label{fig:iras21078_pcomp_vsigma}
\end{figure}

\begin{figure}
\centering
\includegraphics[width=\hsize]{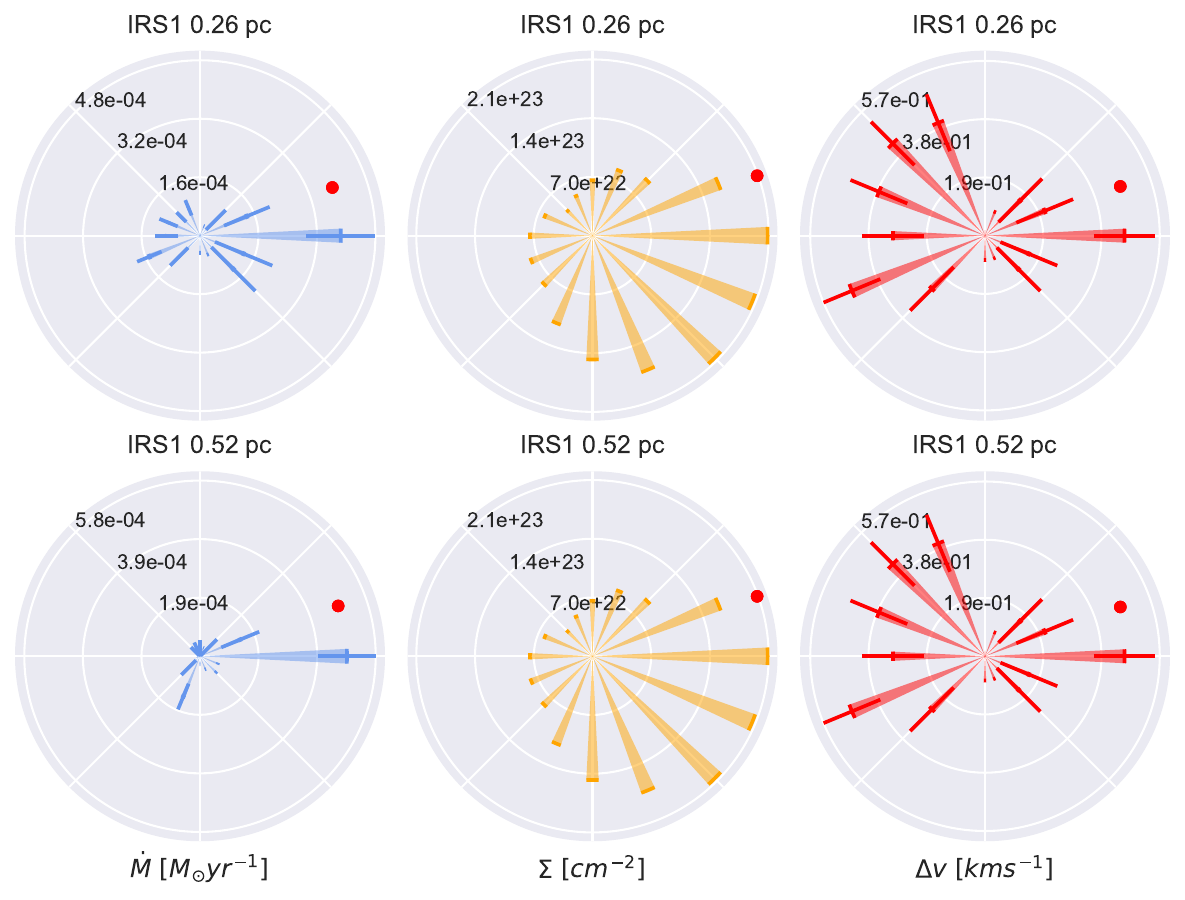}
 \caption{Comparison panels for the clump-centered flow approach. Columns from left to right: Total estimated mass flow rate, surface density and velocity difference for NGC7538 IRS1 (Fig. \ref{fig:dusttags}). Top row: Inner circle at 0.26 pc. Bottom row: Outer circle at 0.52\,pc. The red dots indicate adjacent filaments that are assumed to host the potential flow-direction.}
 \label{fig:ngc7538_pcomp_visgma_irs1}
\end{figure}

As for NGC7538, we can see the diagrams for IRS1, S and IRS9 clumps in Fig. \ref{fig:ngc7538_mf_pol} (for reference, see Fig. \ref{fig:dusttags}). IRS1 shows more extended flow magnitudes towards the west, into the direction of the F-NW filament that declines for increasing distance (about $0.52$ pc), but remains the dominant value. This probably can be explained by a strong velocity difference in the H$^{13}$CO$^+$-data (Fig. \ref{fig:1stmoment_h13co+}). S indicates a more prominent flow-direction towards the northern IRS1, especially for 0.52\,pc distance. It is interesting to note that S seems to be influenced by IRS1 but not vice versa. Assuming that the northern \HII-region may have triggered the formation of IRS1, this region should be slightly older than S. Hence influence of IRS1 on S but not the other way round appears plausible. The mass flow-magnitude towards S also yields no visible influence from the F-SW filament-structure. For IRS9, both distances depict almost the same pattern: the flow rate direction is concentrated to the southern half of the circle with a more obvious contribution from the eastern section than towards the filament connecting IRS9 with S (Fig. \ref{fig:dusttags}).\par

In order to obtain a more precise overview of the contributors to the mass flow rate (Eq. (\ref{eq:mdot})), we conducted the same separation for $\Sigma$ and $\Delta v$ as in section \ref{sec:fila-flows}. We present the comparison-panels for G75 N1 in Fig. \ref{fig:g75_pcomp_vsigma_n-1}, for IRAS21078 CC in Fig. \ref{fig:iras21078_pcomp_vsigma} and for NGC7538 IRS1 in Fig. \ref{fig:ngc7538_pcomp_visgma_irs1}. The diagrams for all other clumps are shown in Appendix \ref{sec:ap-c}.\par

For the most part, the velocity differences appear to have a more dominant impact onto the ``shape'' of the polar magnitude-distribution. For clump G75 N1 (Fig. \ref{fig:g75_pcomp_vsigma_n-1}) however, the surface density seems to be the strongest contributor. We can for example see that the strong flow rates in G75 S1 are dominated by the velocity difference towards S2 (Fig. \ref{fig:g75_pcomp_vsigma_s-1}), while the surface density shows orientation towards the east and does not align at all with the expected filament direction (red dot) and only to a small degree with the velocity differences. The same holds for S2 towards S1 (Fig. \ref{fig:g75_pcomp_vsigma_s-2}). IRAS21078 CC exhibits one spike into F-NW direction that clearly stems from the velocity difference for close distance (Fig. \ref{fig:iras21078_pcomp_vsigma}). The same holds for the outer annulus, where the surface density now points into F-NW direction, but the velocity difference towards the south-east is stronger and thus affects the overall mass flow to a larger degree. This also is visible for NGC7538S (Fig. \ref{fig:ngc7538_pcomp_visgma_s}).\par

\begin{figure}[t!]
    \centering
    \includegraphics[width=1\linewidth]{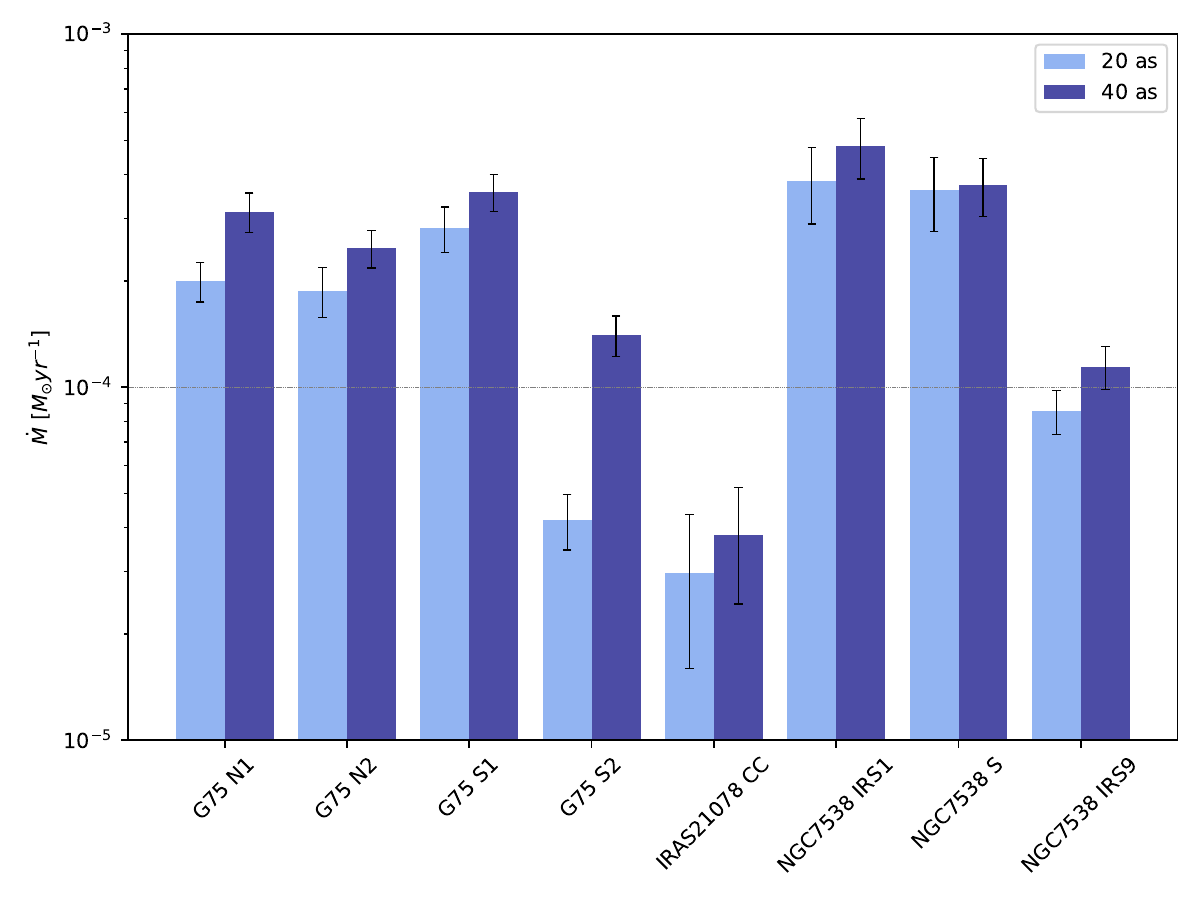}
    \caption{Peak-values of the estimated mass flow rate towards clumps. Light blue bars show the values for \SI{20}{\arcsecond} distance while dark blue bars show values for \SI{40}{\arcsecond} distance. The error bars show the maximum error of the respective mass-flow rate. Notice that this chart shows the maximum values of $\dot{M}$ alone, and contains no information about the direction from the polar plots.} 
    \label{fig:polar_mf_max_barchart}
\end{figure}

As we summarize in Fig. \ref{fig:polar_mf_max_barchart}, the peak-values for all of the mass flow rates occur in the outer annulus of \SI{40}{\arcsecond}. This indicates a larger fraction of mass being transported towards the clumps on the larger scale.

\section{Discussion, conclusions and summary}\label{sec:conclusion_summary}

In this paper, we analyzed the kinematics of molecular gas in three high-mass star-forming molecular clouds. In order to do this, we estimated the mass flow rates and divided the analysis into mass flows along filamentary structures and flows onto clumps. Both approaches indicate that mass is transported along these structures. Along filamentary cloud-structures, we notice increasing mass flow magnitudes close to denser regions. The peak values are in the order of $10^{-3}\,M_{\odot}$ for G75 and $10^{-5}\,M_{\odot}$ and $10^{-4}\,M_{\odot}$ for IRAS21078 and NGC7538 respectively. As given in Table \ref{tab:clumps}, we note that IRAS21078 is a lower mass clump which therefore exhibits the lowest overall mass flow rate.\par 

The analysis of the flows onto clumps focused on the interpretation of polar diagrams, that show the orientation of the mass flow rate magnitude towards a clump. In G75, the direction of strongest mass flow rate indicates a potential feeding of the northern half of the cloud through the central connecting filament-structure. Similar patterns appear in NGC7538 for the two clumps IRS1 and S. As for IRAS21078, the distribution of the flow rates indicates infall from the north-western filament-structure, but one has to bear in mind the comparably large errors; it however appears possible that infall of material through these channels occurs, as it already has been subjected to studies with similar findings on smaller spatial scales.\par

The separation of the two main contributors to the mass flow rates into velocity difference and surface density showed interesting patterns for both methods: as for the flows along the G75 CF-structure, the surface density dictates the overall shape of the curve; spikes are on the other hand contributed by steep velocity differences. Sharp increases in velocity differences often affect the mass flow curve to a large degree, and the data indicate that they often stem from feedback from the \HII-regions. In the case of the polar diagrams, the orientation of the surface density towards filament-structures was sometimes not visible due to a strong velocity difference into a different direction. There are cases however, where the two factors of the mass flow rate equation are more aligned. This is overall very interesting, since an orientated distribution of dust for close proximity might hint at channel- or filament-structures that serve as a gas-funnel. The fact that the larger velocity differences sometimes do not point toward adjacent clump- or filament direction might be caused by expanding shells from nearby \HII-regions; this very likely is the case for G75 S1 and S2. As for the clumps were we also spot velocity differences into filament-directions, we can assume that it potentially is accretionally-driven through the filaments. For the regions were this behavior does not occur, we can tentatively infer that the \HII-regions are responsible due to feedback-processes: this assessment is supported by the fact that all clumps that show flow rate magnitudes in unexpected directions in the polar plots are located close to a \HII-region. 

Similar to the assertion of a dominant contributor in the mass flow rate, large flow-magnitude values close to the edge of a cloud (cf. Fig. \ref{fig:ngc7538_mf_pol}) do not necessarily imply increasing dynamics in the gas flow, but could just be a consequence of the steep gradient in the surface density profile (cf. Eq. (\ref{eq:mdot}) and Fig. \ref{fig:dusttags}, especially in the vicinity of IRS1 and IRS9 in NGC7538 that are located close to the edge of the 1.2\,mm dust emission).\par

If we assume that the accretion is mainly gravitationally-driven close to denser regions (e.g. clumps), the material should fall onto the clumps; driving mechanisms like turbulence on larger scales have been numerically studied by e.g. \citealt{Padoan}. Large-scale infall would predecess fragmentation of the cloud and further collapse of proto-stars that exceed their individual Jeans mass (\citealt{Vazquez}). In the case of the here presented regions, we found strong indications for feedback processes to be the driving force behind gas dynamics in the clouds. The regions with \HII-regions can differ in their behavior, and the plots indicate that they certainly do. Their influence is strongly noticeable in disrupted flow-behavior and the lack of molecular line data (Fig. \ref{fig:dusttags}). For this reason, especially the north-western part of NGC7538 has to be excluded from the accreting-flow assessment in general due to the \HII-region (e.g. \citealt{Beuther2022}) (see also Sect. \ref{sec:results} and Fig. \ref{fig:dusttags}). Filaments appear at the boundaries of \HII-regions in all presented sources with strong indications for enhanced mass flow. This supports the simulations predicting filament formation at the crossing of shocks (e.g. \citealt{PudritzKevlahan2013}). We did only include the analysis for the HCO$^+$ and H$^{13}$CO${^+}$-molecules in the main section, since the other molecules that exhibit a similar spatial distribution (e.g. HCN and H$_2$CO) did not yield additional information in their apparent kinematics.

\begin{acknowledgements}
    This work is based on observations carried out under project number 121-20 with the IRAM 30m telescope. IRAM is supported by INSU/CNRS (France), MPG (Germany) and IGN (Spain).\\
    
    A.P. acknowledges financial support from the UNAM-PAPIIT IG100223 grant, the Sistema Nacional de Investigadores of SECIHTI, and from the SECIHTI project number 86372 of the `Ciencia de Frontera 2019’ program, entitled `Citlalc\'oatl: A multiscale study at the new frontier of the formation and early evolution of stars and planetary systems’, M\'exico.\\

    S.F. acknowledges support from the National Key R\&D program of China grant (2025YFE0108200), National Science Foundation of China (12373023,  1213308).
\end{acknowledgements}

\bibliographystyle{aa}
\bibliography{Refs}

\begin{appendix}
\section{Additional mass flow and comparison graphs (along filaments)}\label{sec:ap-a}

\begin{figure}[htbp!]
    \centering
    \includegraphics[width=\hsize]{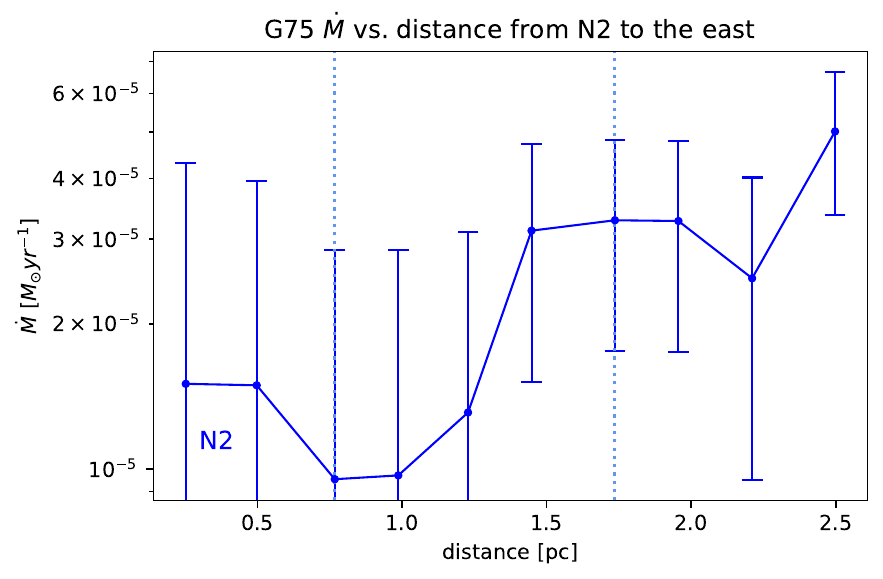}
    \includegraphics[width=\hsize]{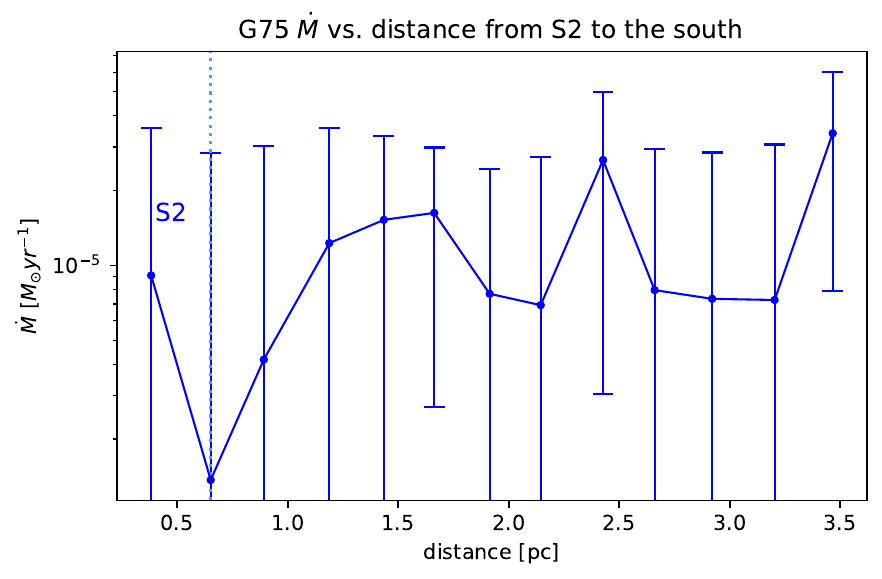}
    \caption[mass flow rate G75 N2]{{\bf Top:} Estimated mass flow rate as a function of the distance from N2 eastwards in G75 (see Fig. \ref{fig:dusttags}). The blue graph follows the west-east direction starting at N2 from left to right. The dotted blue lines mark the approximate positions of the density enhancements along the structure. {\bf Bottom:} The same for S2.}
    \label{fig:g75_mf_north}
\end{figure}

\begin{figure}[htbp!]
\centering
\includegraphics[width=\hsize]{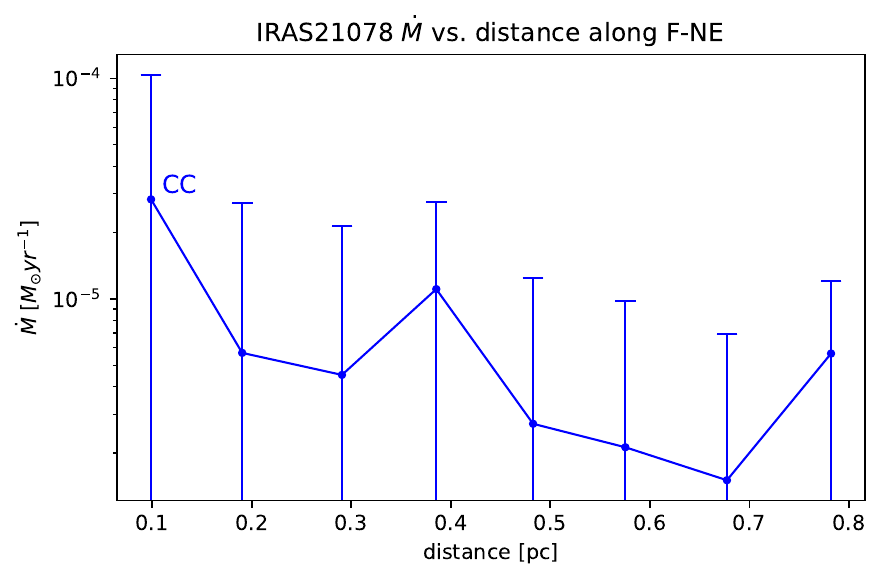}
\includegraphics[width=\hsize]{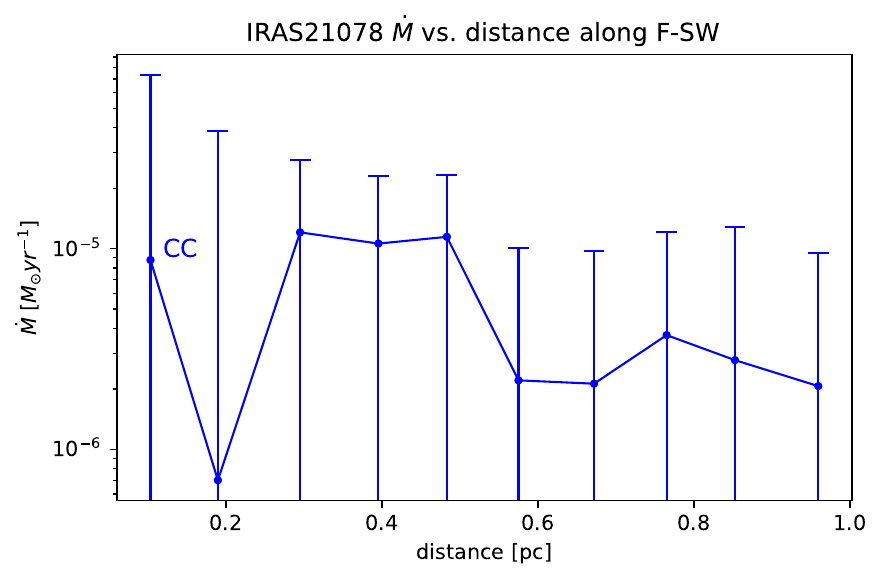}
\caption{{\bf Top:} Mass flow estimates as a function of distance (of adjacent points) along the F-NE filament-structure in IRAS21078 (cf. Fig. \ref{fig:dusttags}). The direction of measurement is from the clump outwards, where the ``CC'' annotation indicates the starting point. {\bf Bottom:} The same for F-SW.}
 \label{fig:iras_fil_ne-f}
\end{figure}

\begin{figure}[htbp!]
\centering
\includegraphics[width=\hsize]{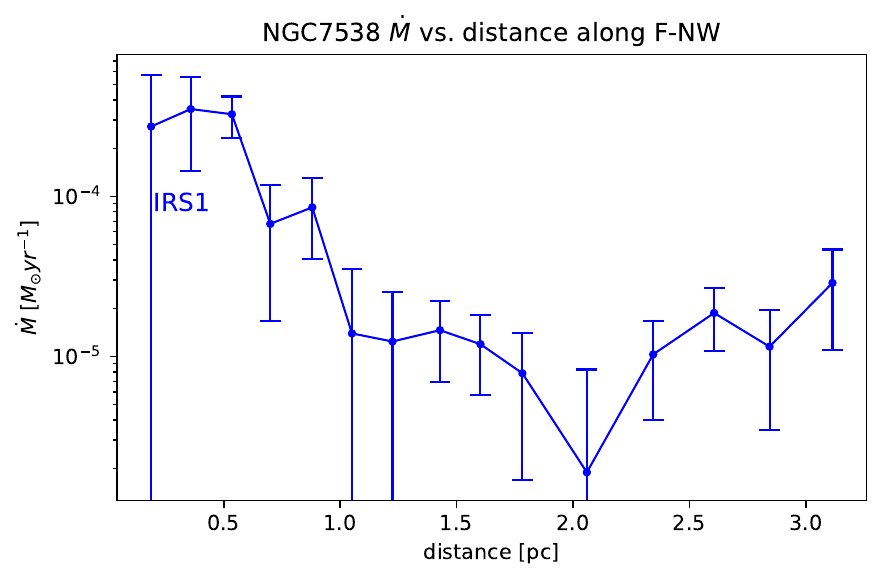}
 \caption{Mass flow estimates as a function of distance (of adjacent points) along the F-NW filament-structure in NGC7538 (cf. Fig. \ref{fig:dusttags}).}
 \label{fig:ngc_fil_f-nw}
\end{figure}

\begin{figure*}
\centering
\includegraphics[width=\hsize]{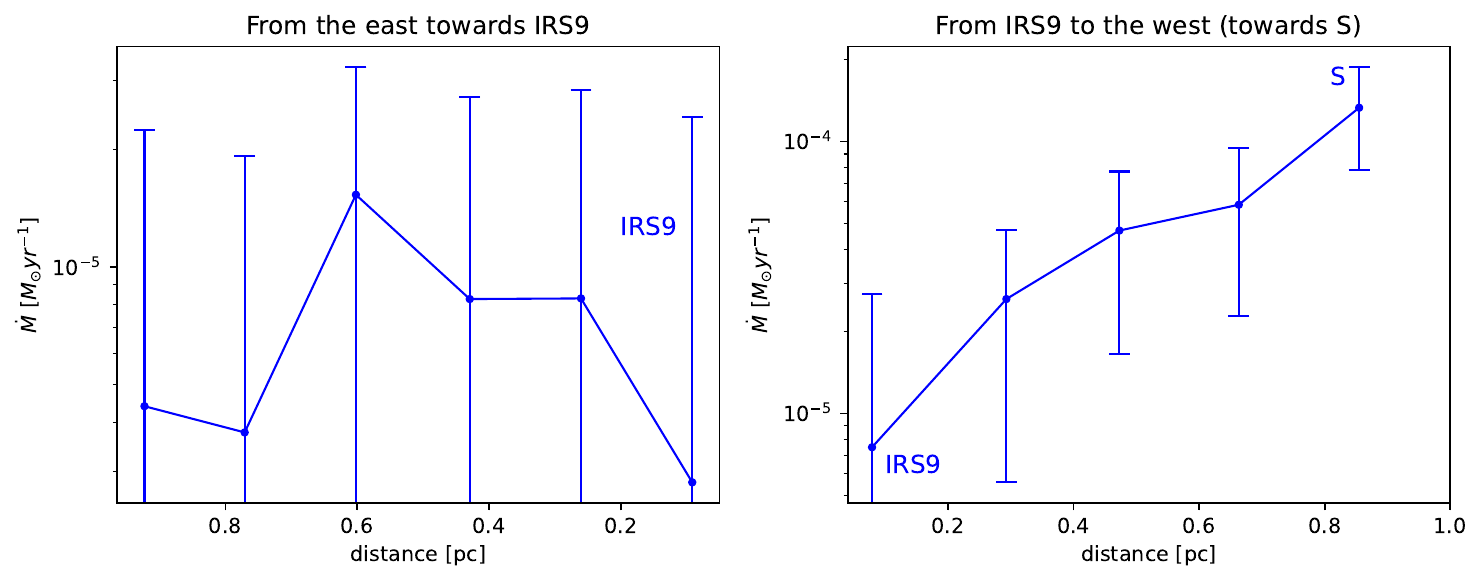}
 \caption{mass flow estimates as a function of distance (of adjacent points) along the F-E filament-structure in NGC7538. The panels are aligned such that the direction of measurement can be traced directly from Fig. \ref{fig:dusttags}.}
 \label{fig:ngc_fil_f-e}
\end{figure*}

\begin{figure*}[htbp!]
\centering
\includegraphics[width=0.9\hsize]{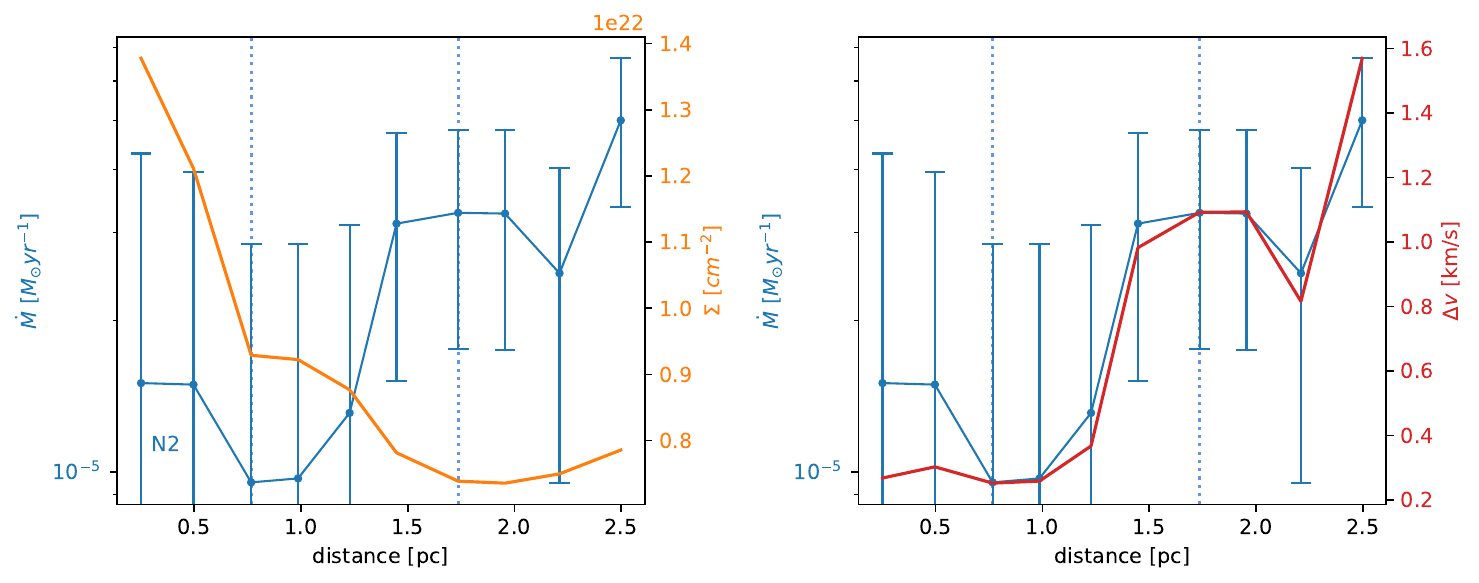}
 \caption[Comparison of contributors to $\dot{M}$ in G75 from N2 to the east]{Light blue graph: mass flow rate as a function of the distance between adjacent points in G75 from N2 eastwards (Fig. \ref{fig:dusttags}). Orange graph in the left panel: surface density in G75 from N2 eastwards. Red graph in the right panel: velocity differences in G75 from N2 eastwards. The dotted blue lines mark the approximate positions of the density enhancements along the structure.}
 \label{Fig:g75_n2_comp_mfcoldens}
\end{figure*}

\begin{figure*}[htbp!]
\centering
\includegraphics[width=0.9\hsize]{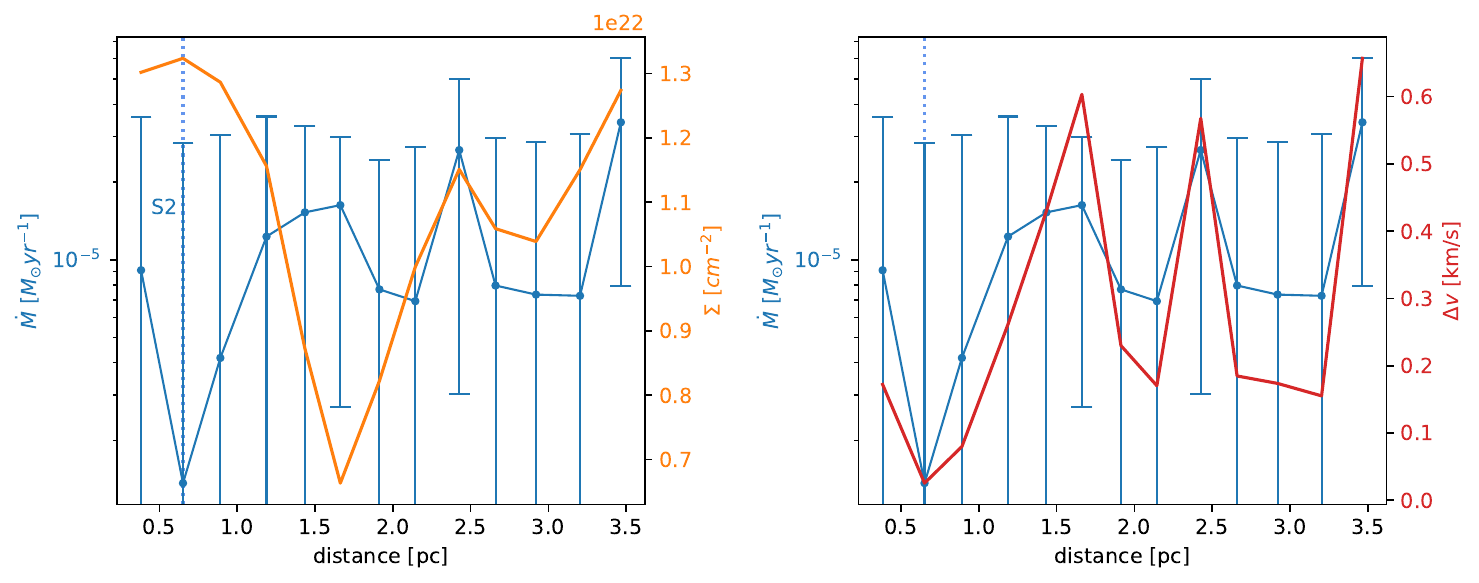}
 \caption[Comparison of contributors to $\dot{M}$ in G75 from S2 to the south]{As for Fig. \ref{Fig:g75_n2_comp_mfcoldens} but for S2.}
 \label{Fig:g75_s2_comp_mfcoldens}
\end{figure*}

\begin{figure*}[htbp!]
\centering
\includegraphics[width=0.9\hsize]{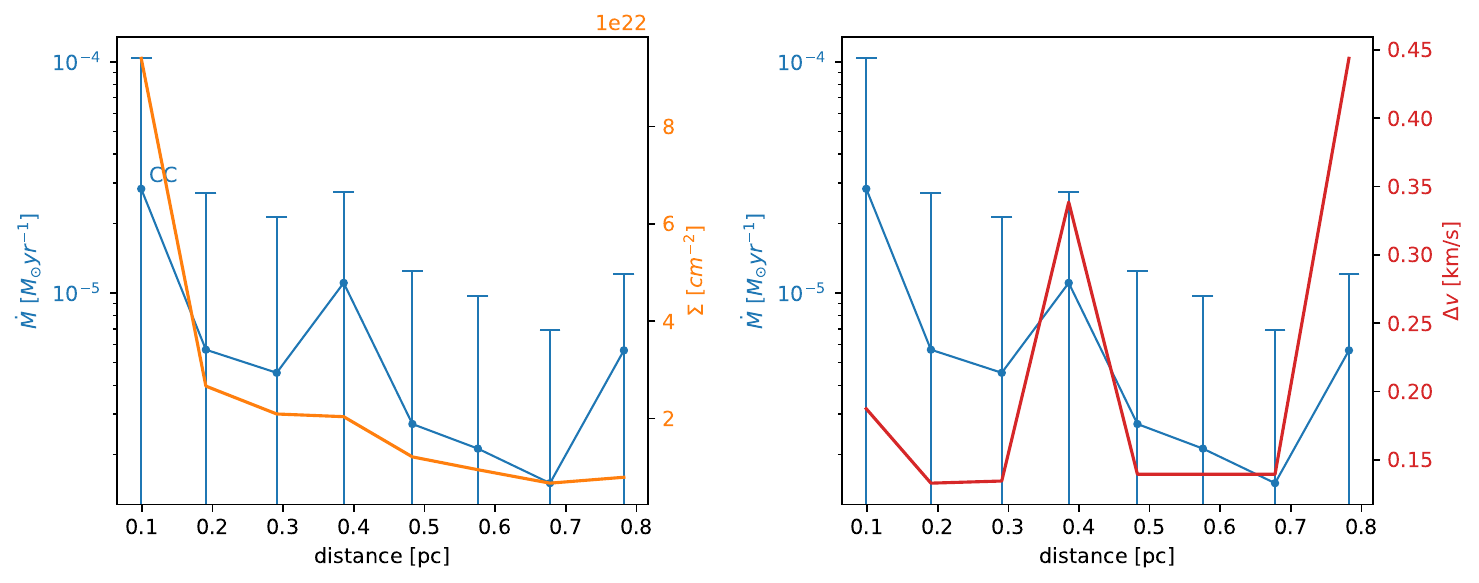}
 \caption[Comparison of contributors to $\dot{M}$ in IRAS21078 F-NE]{Light blue graph: mass flow rate as a function of the distance between adjacent points in IRAS21078 F-NE (Fig. \ref{fig:dusttags}). Orange graph in the left panel: surface density along the F-NE-structure in IRAS21078. Red graph in the right panel: velocity differences along the F-NE-structure in IRAS21078.}
 \label{Fig:iras21078_fne_comp_mfcoldens}
\end{figure*}

\begin{figure*}[htbp!]
\centering
\includegraphics[width=0.9\hsize]{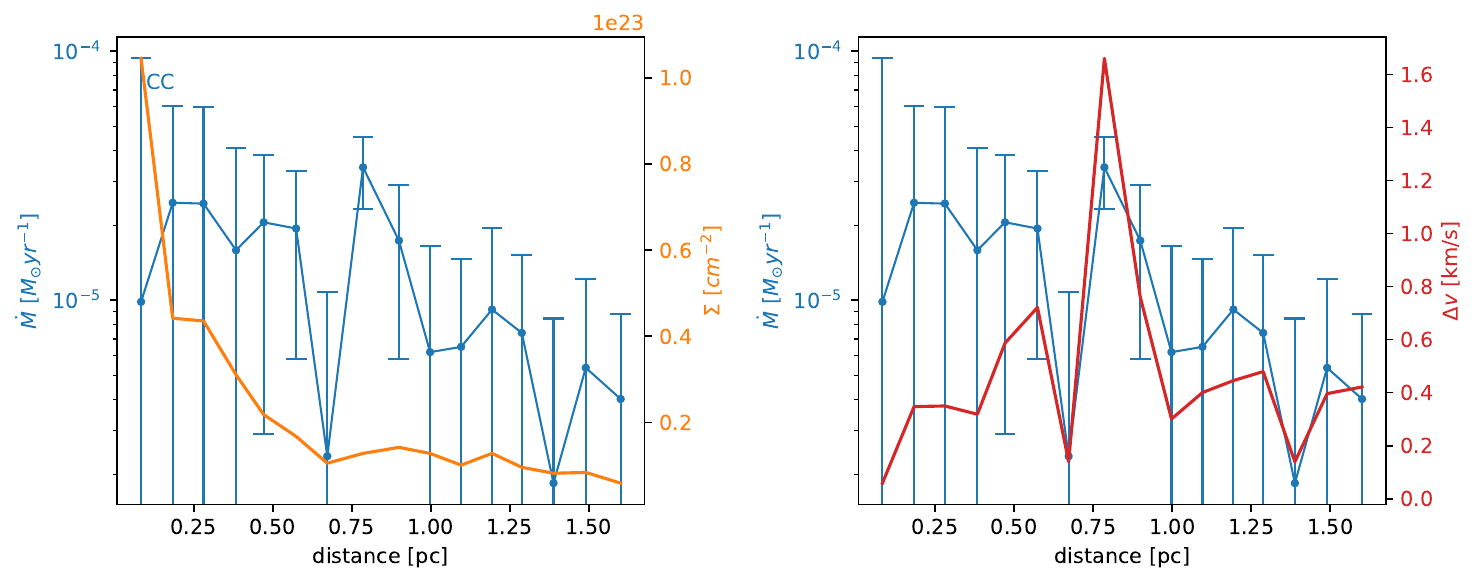}
 \caption[Comparison of contributors to $\dot{M}$ in IRAS21078 F-NW]{As for Fig. \ref{Fig:iras21078_fne_comp_mfcoldens} but for IRAS21078 F-NW.}
 \label{Fig:iras21078_fnw_comp_mfcoldens}
\end{figure*}

\begin{figure*}[htbp!]
\centering
\includegraphics[width=0.9\hsize]{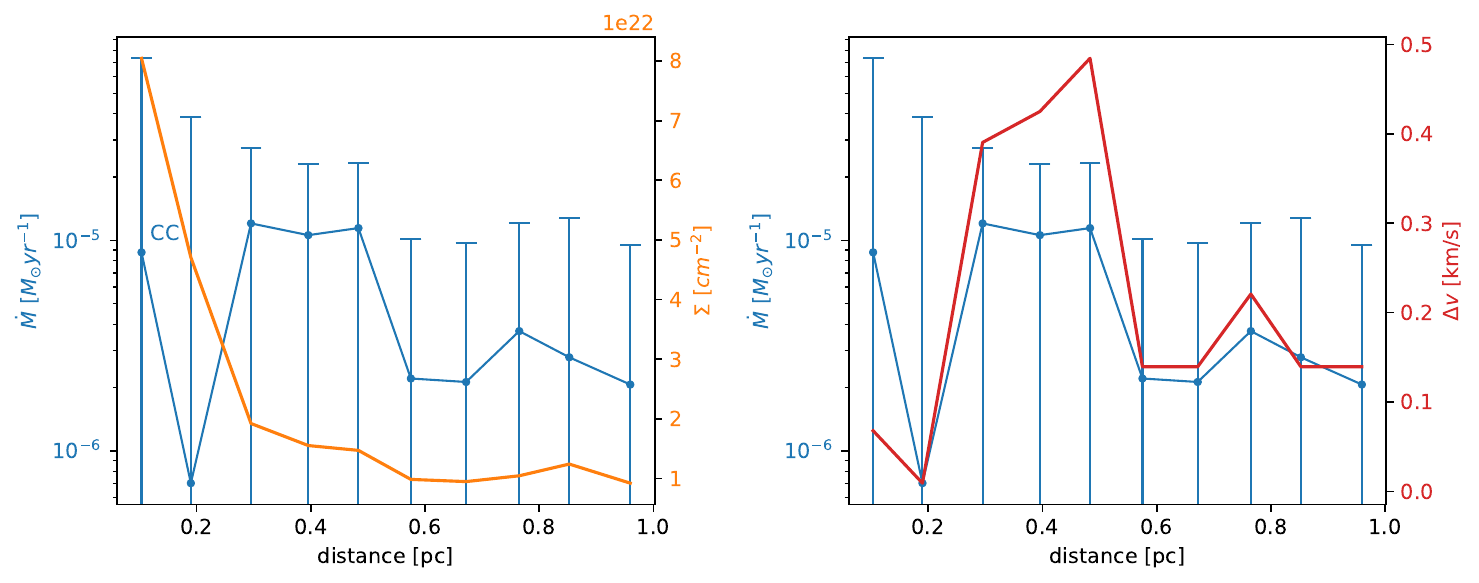}
 \caption[Comparison of contributors to $\dot{M}$ in IRAS21078 F-SW]{As for Fig. \ref{Fig:iras21078_fne_comp_mfcoldens} but for IRAS21078 F-SW.}
 \label{Fig:iras21078_fsw_comp_mfcoldens}
\end{figure*}

\begin{figure*}[htbp!]
\centering
\includegraphics[width=0.9\hsize]{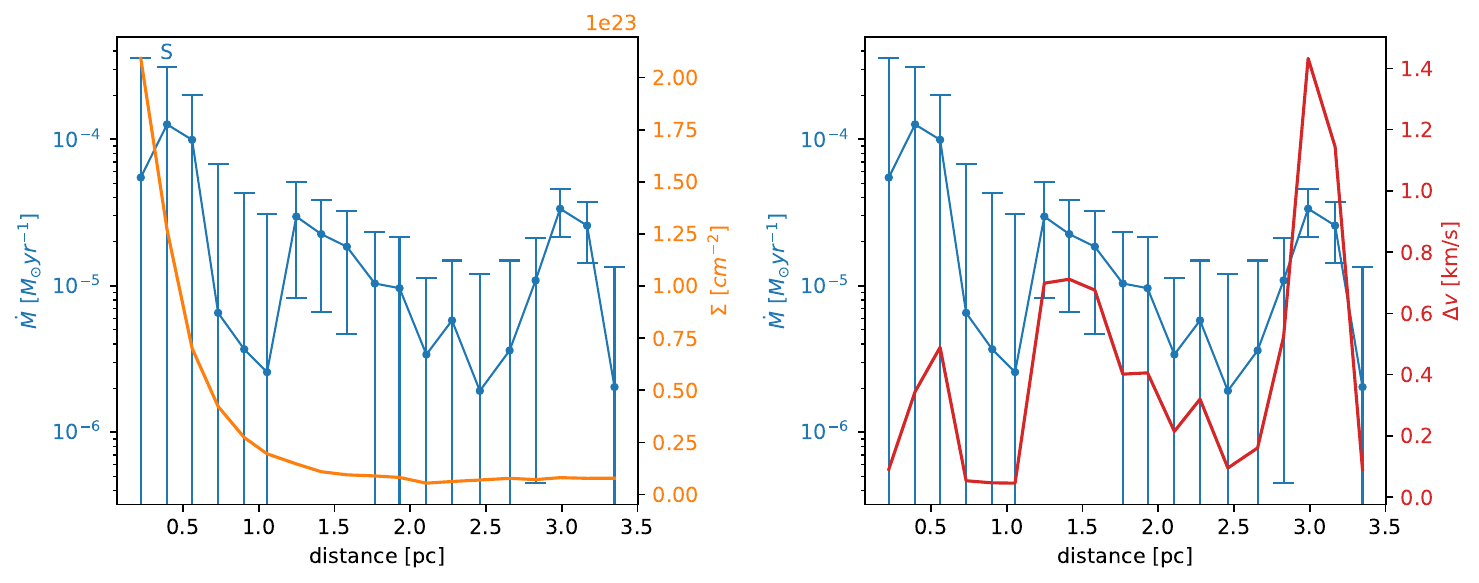}
 \caption[Comparison of contributors to $\dot{M}$ in NGC7538 F-SW]{Light blue graph: mass flow rate as a function of the distance between adjacent points in NGC7538 F-SW (Fig. \ref{fig:dusttags}). Orange graph in the left panel: surface density along the F-SW-structure in NGC7538. Red graph in the right panel: velocity differences along the F-SW-structure in NGC7538.}
 \label{Fig:ngc7538_fsw_comp_mfcoldens}
\end{figure*}

\begin{figure*}[htbp!]
\centering
\includegraphics[width=0.9\hsize]{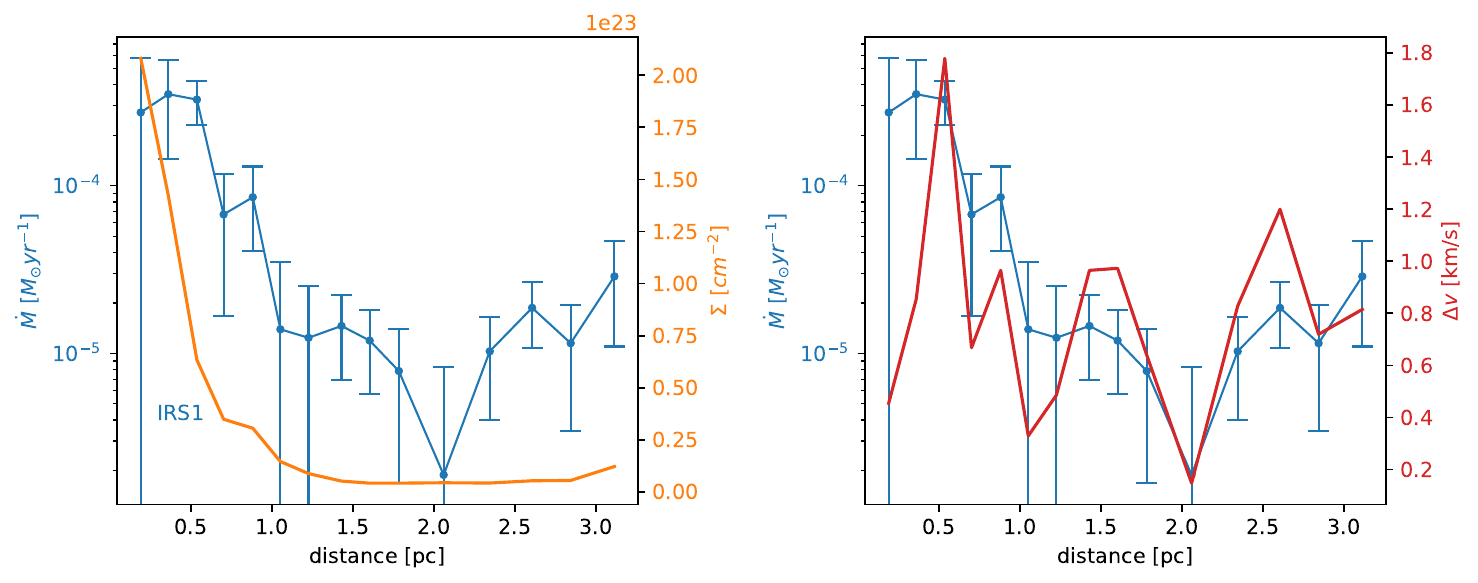}
 \caption[Comparison of contributors to $\dot{M}$ in NGC7538 F-NW]{As for Fig. \ref{Fig:ngc7538_fsw_comp_mfcoldens} but for NGC7538 F-NW.}
 \label{Fig:ngc7538_fnw_comp_mfcoldens}
\end{figure*}

\begin{figure*}[htbp!]
\centering
\includegraphics[width=0.9\hsize]{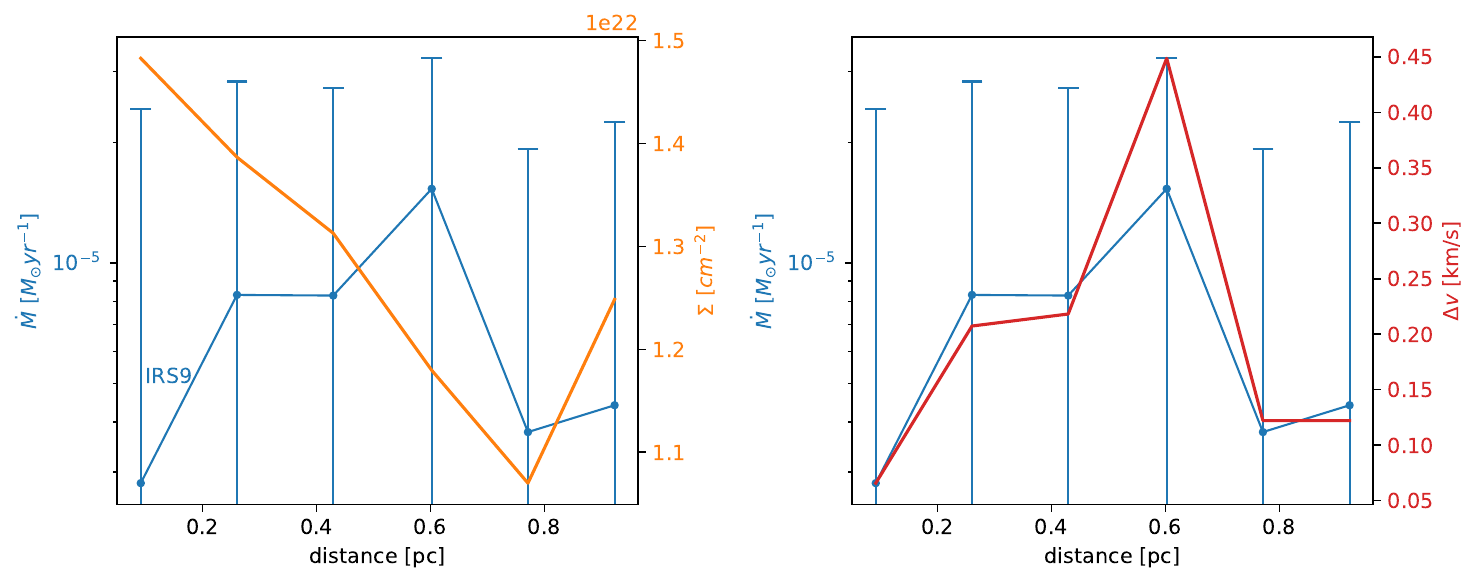}
 \caption[Comparison of contributors to $\dot{M}$ along NGC7538 F-E (from IRS9 to the east)]{As for Fig. \ref{Fig:ngc7538_fsw_comp_mfcoldens} but for NGC7538 F-E (from IRS9 towards the east).}
 \label{Fig:ngc7538_fee_comp_mfcoldens}
\end{figure*}

\begin{figure*}[htbp!]
\centering
\includegraphics[width=0.9\hsize]{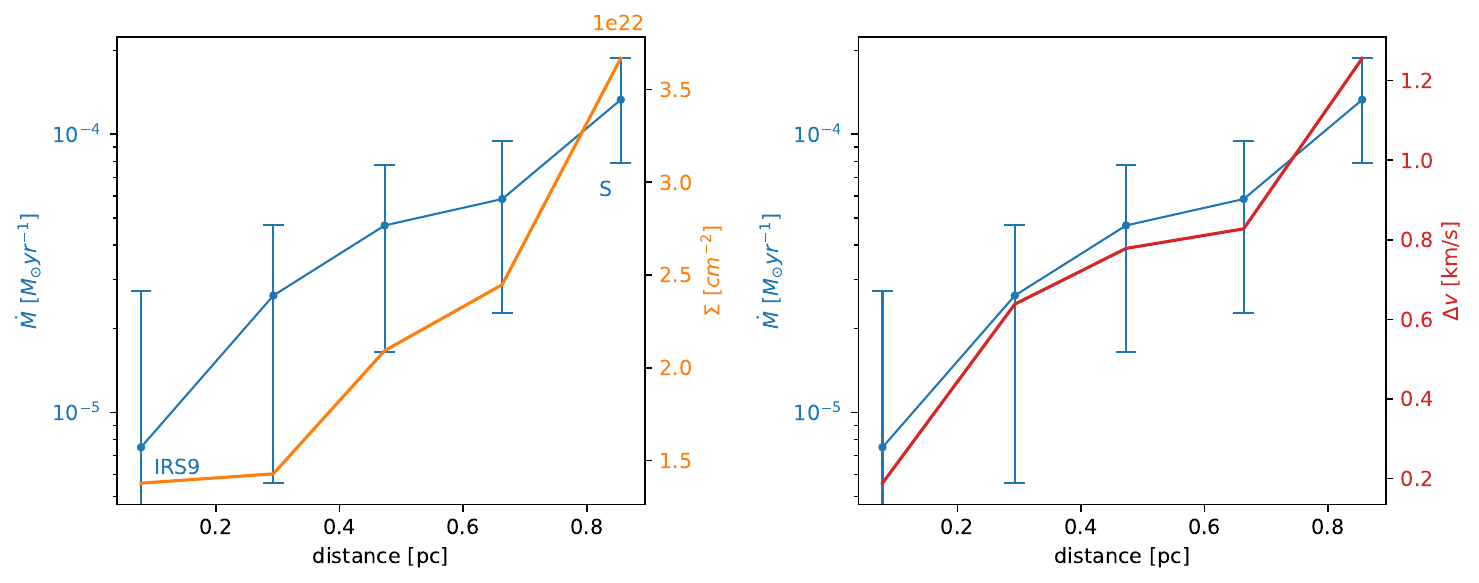}
 \caption[Comparison of contributors to $\dot{M}$ along NGC7538 F-E from IRS9 towards S]{As for Fig. \ref{Fig:ngc7538_fee_comp_mfcoldens} but from NGC7538 IRS9 towards S.}
 \label{Fig:ngc7538_few_comp_mfcoldens}
\end{figure*}

\clearpage
\section{Additional comparison-graphs (polar diagrams)}\label{sec:ap-c}

\begin{figure}[htb!]
\centering
\includegraphics[width=1\linewidth]{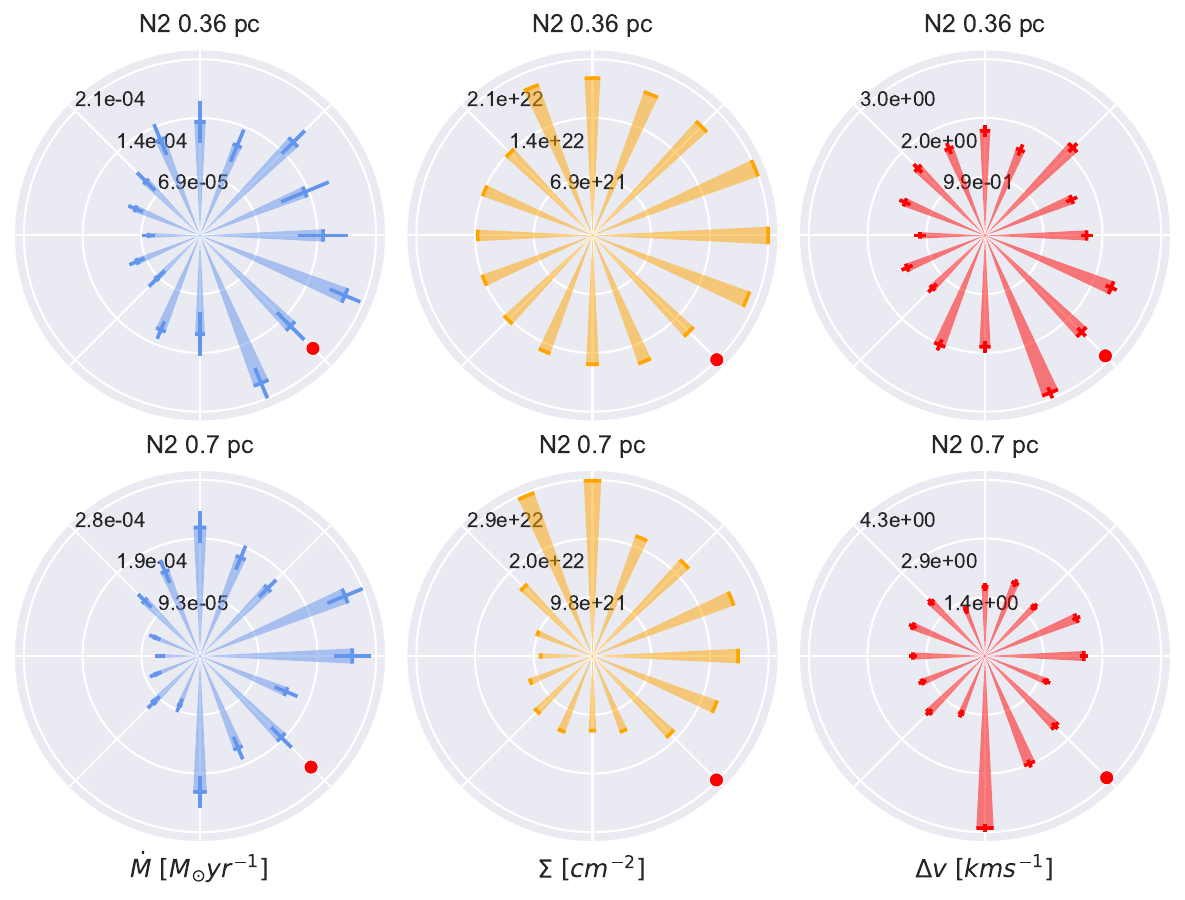}
 \caption{Comparison panels for the clump-centered flow approach. Columns from left to right: Total estimated mass flow rate, surface density and velocity difference for G75 N2 (Fig. \ref{fig:dusttags}). The red dots indicate adjacent filaments which are assumed to host the potential flow-direction.}
 \label{fig:g75_pcomp_vsigma_n-2}
\end{figure}

\begin{figure}[htb!]
\centering
\includegraphics[width=\linewidth]{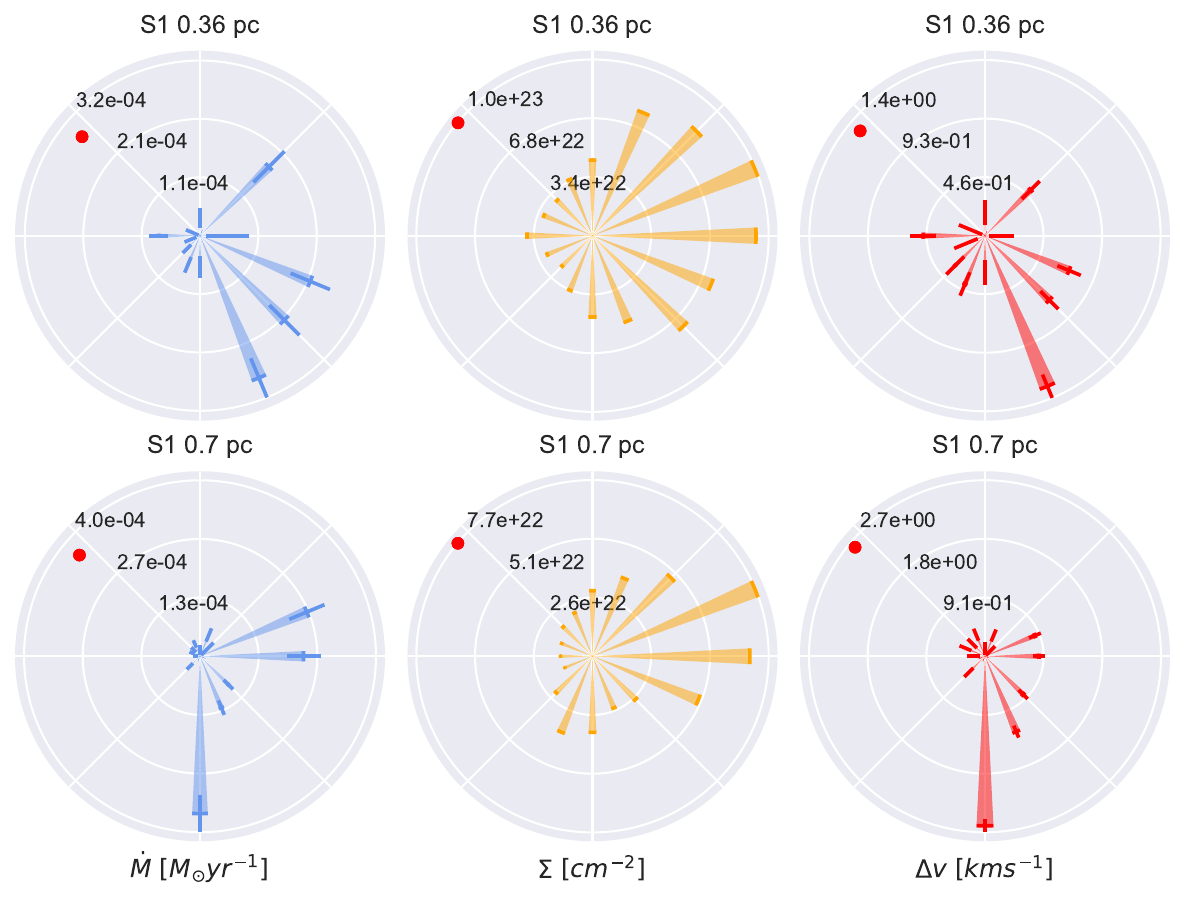}
 \caption{As for Fig. \ref{fig:g75_pcomp_vsigma_n-2} but for G75 S1.}
 \label{fig:g75_pcomp_vsigma_s-1}
\end{figure}

\begin{figure}[htb!]
\centering
\includegraphics[width=\linewidth]{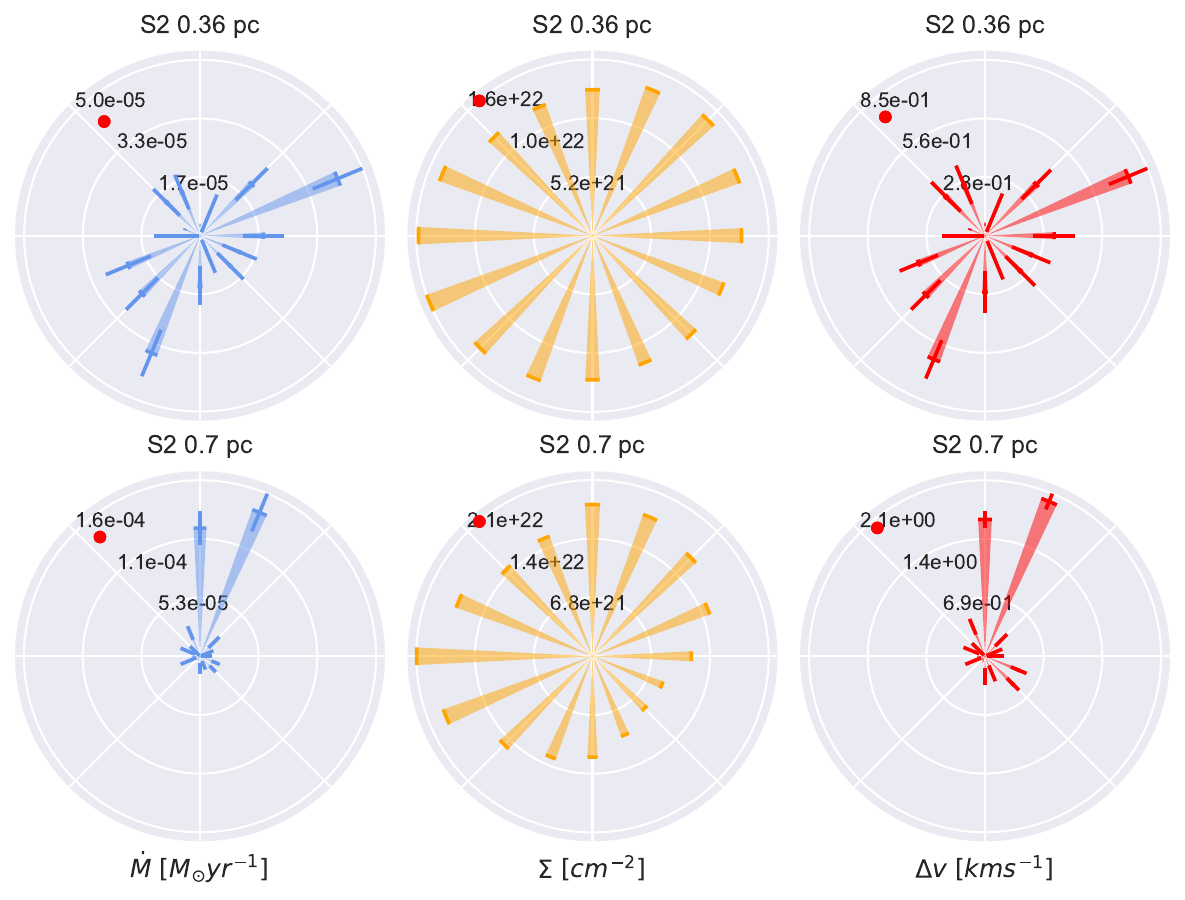}
 \caption{As for Fig. \ref{fig:g75_pcomp_vsigma_n-2} but for G75 S2.}
 \label{fig:g75_pcomp_vsigma_s-2}
\end{figure}

\begin{figure}[htb!]
\centering
\includegraphics[width=\linewidth]{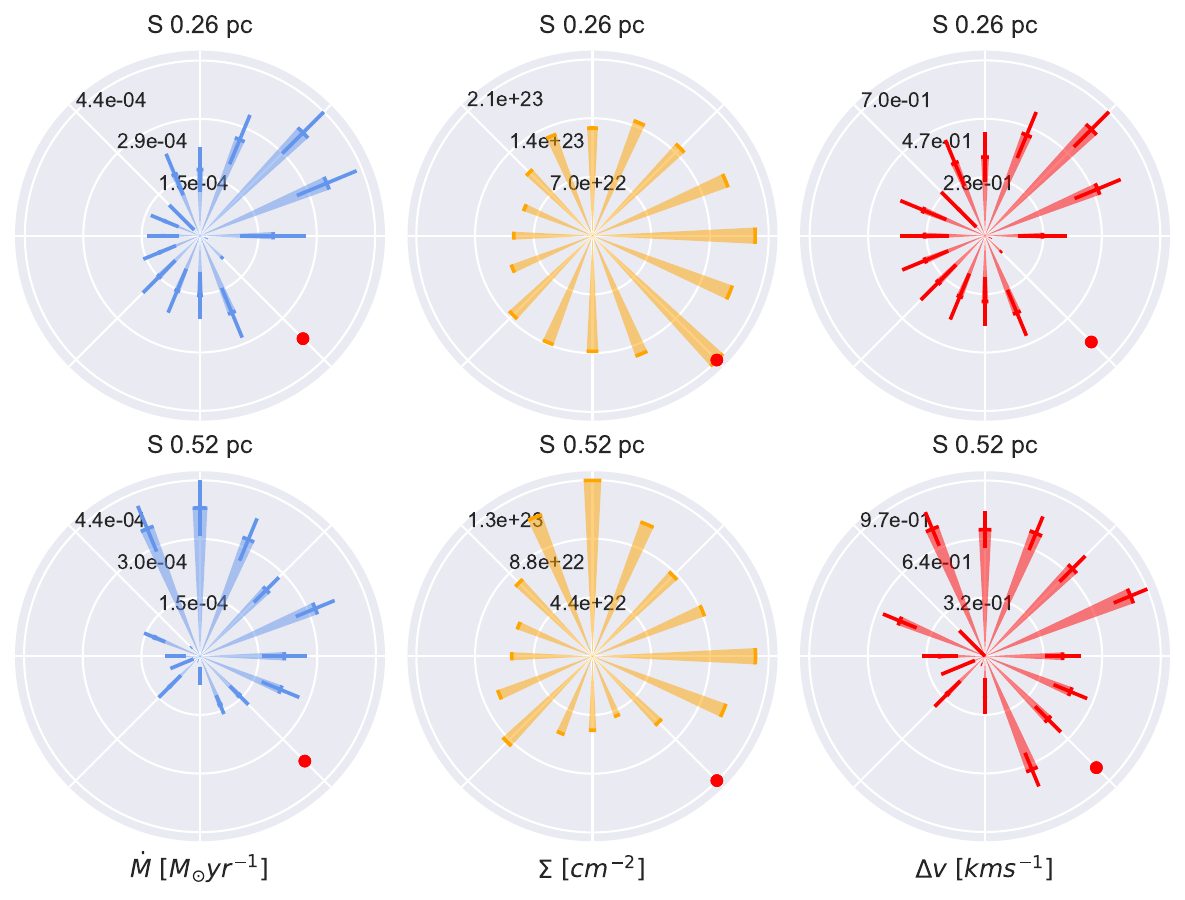}
 \caption{As for Fig. \ref{fig:g75_pcomp_vsigma_n-2} but for NGC7538 S.}
 \label{fig:ngc7538_pcomp_visgma_s}
\end{figure}

\begin{figure}[htb!]
\centering
\includegraphics[width=\linewidth]{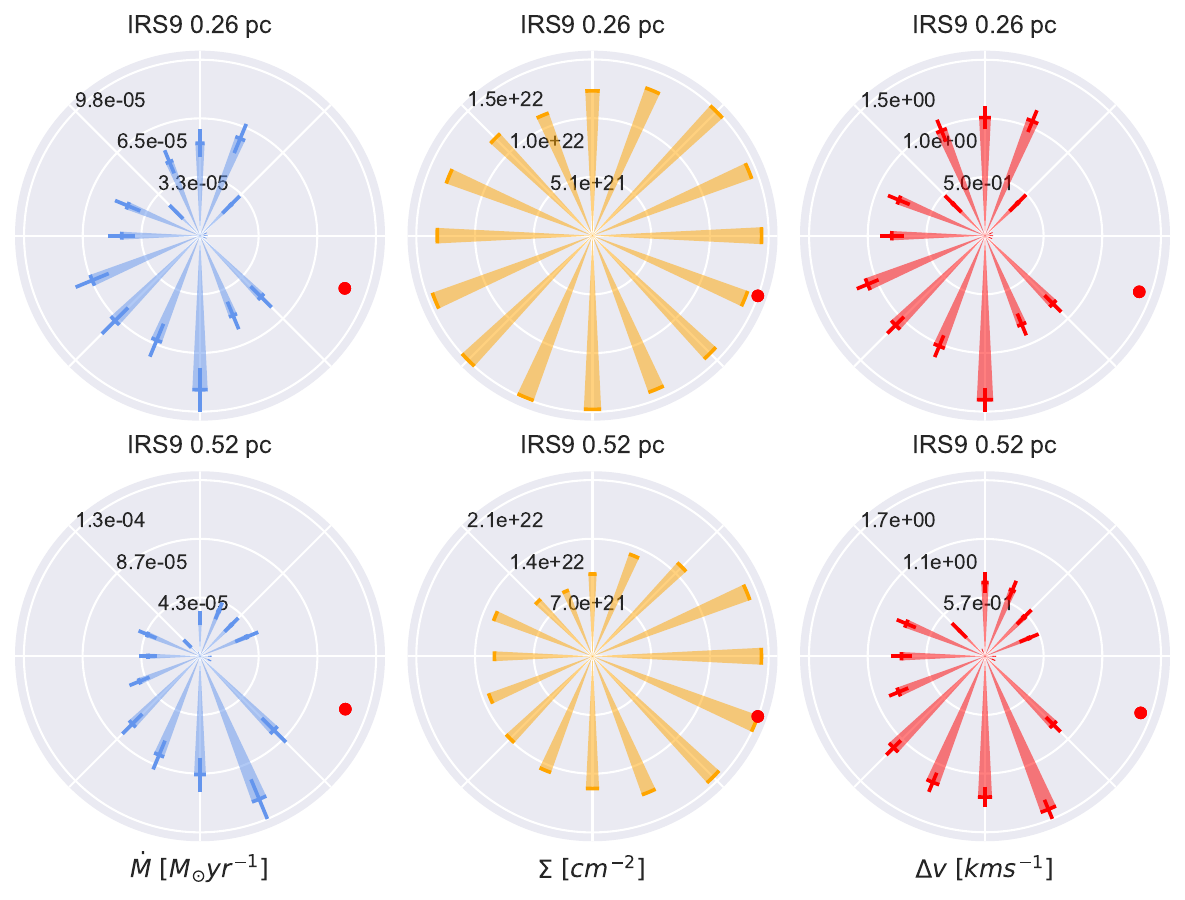}
 \caption{As for Fig. \ref{fig:g75_pcomp_vsigma_n-2} but for NGC7538 IRS9.}
 \label{fig:ngc7538_pcomp_visgma_irs9}
\end{figure}

\clearpage
\onecolumn
\section{Moment zero maps and emission spectra}\label{sec:ap-d}
\begin{figure}[htb!]
    \centering
    \includegraphics[width=0.9\linewidth ,keepaspectratio] {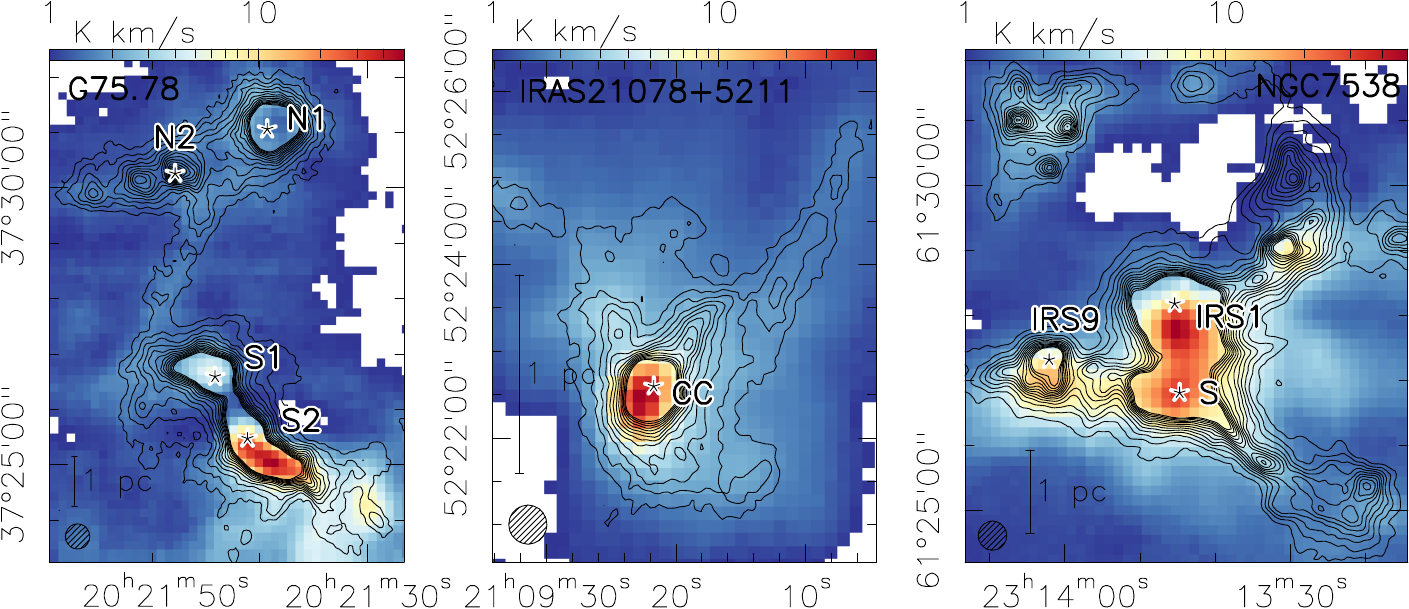}
    \includegraphics[width=0.9\linewidth ,keepaspectratio] {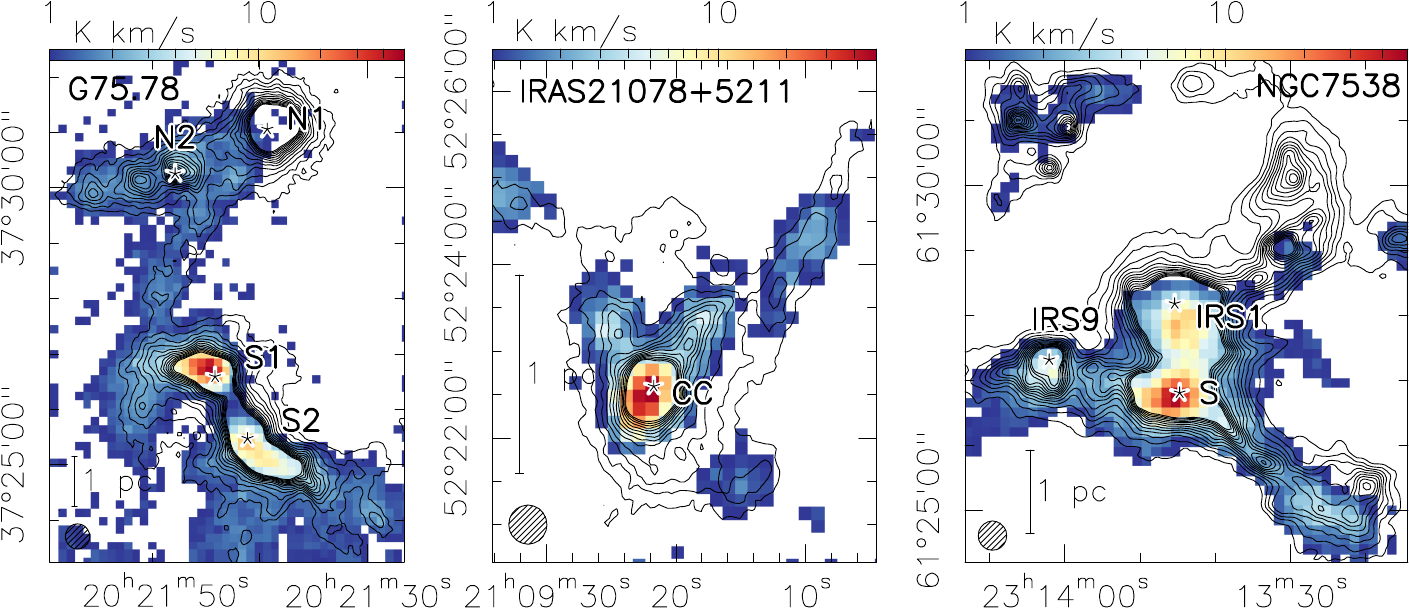}
    \caption[Moment zero maps HCO$^+$\,($1-0$)]{\textbf{Top:} Moment zero (integrated intensity) maps for the HCO$^+$\,($1-0$)-line. The beam size is \SI{27}{\arcsecond} (lower left corner). The scale bar indicates the length of 1 parsec in each panel. Contour lines show the 1.2\,mm dust continuum ranging from 3$\sigma$ to 39$\sigma$ in 3$\sigma$-steps. \textbf{Bottom:} The same for H$^{13}$CO$^+$\,($1-0$).}
    \label{fig:momzero_hco+}
\end{figure}

\begin{figure}[htb!]
    \centering
    \includegraphics[width=0.8\linewidth ,keepaspectratio]{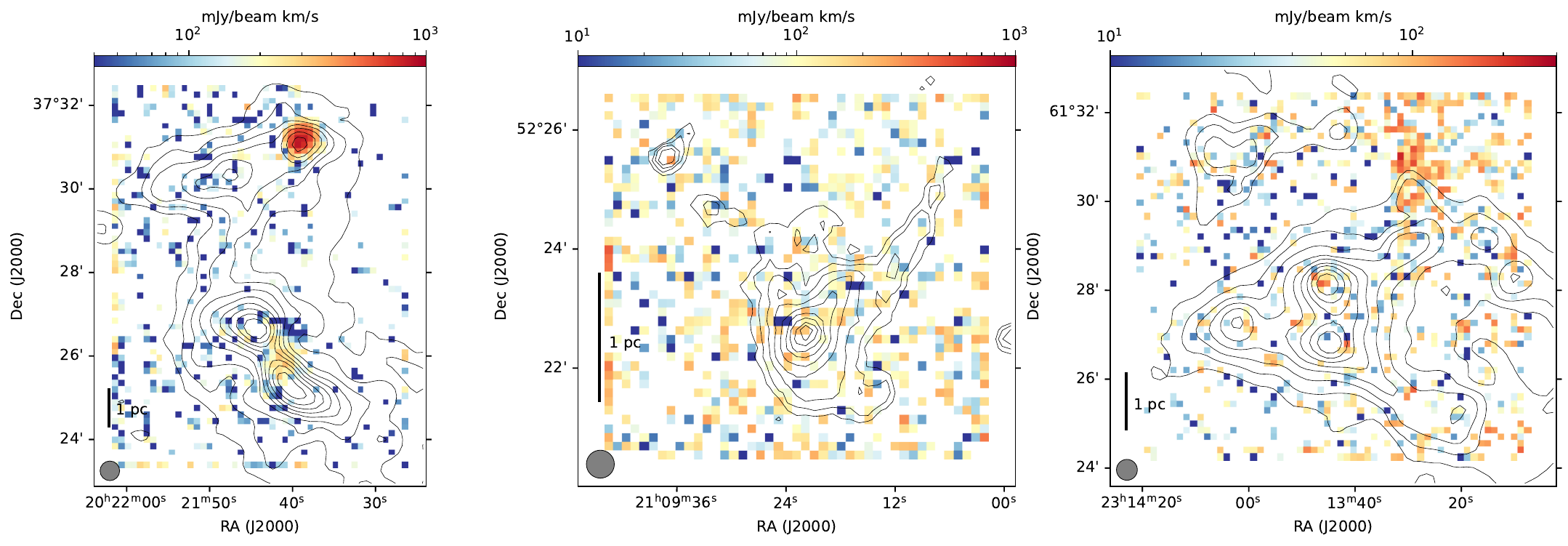}
    \caption[Free-free contribution]{Moment zero maps of the H41$\alpha$ recombination line, tracing free-free contribution. Black contours outline the free-free corrected 1.2\,mm dust continuum (except for IRAS21078) on log-scale starting from 5$\sigma$ (except for IRAS21078 starting from 3$\sigma$) to peak value for reference. While for G75 (left panel) and NGC7538 (right panel) the strongest emission fits the contours in the lower panels of Fig. \ref{fig:dusttags}, the middle panel for IRAS21078 shows only noise and no significant contribution from the \HII-region outlined in NVSS. The edges have been set to values of zero in order to get rid of stripe artifacts from the OTF scanning process. All blank spots represent zero values due to the log-scaling.}
\end{figure}
\twocolumn

\begin{figure}[htb!]
    \centering
    \includegraphics[width=0.4\linewidth]{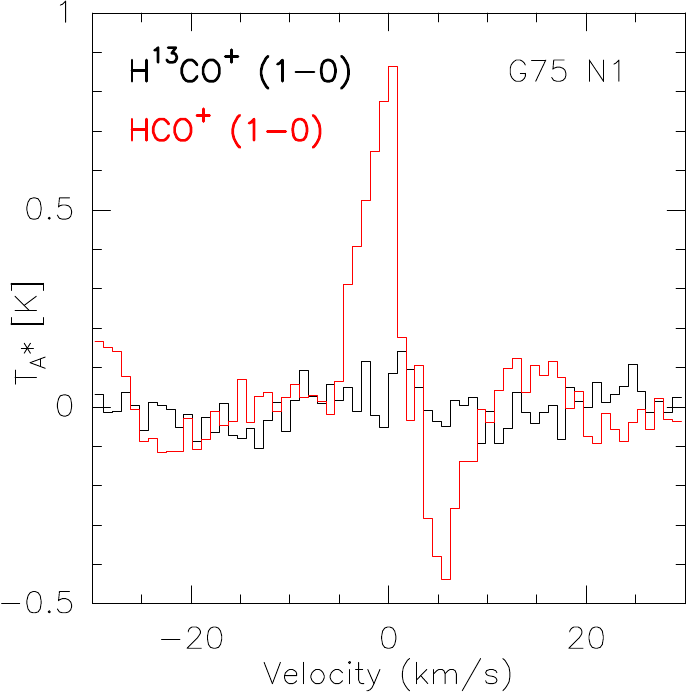}
    \caption{Molecular emission in G75 N1. Due to the lack of H$^{13}$CO$^+$ emission close to N1, we widened the search radius from \SI{10}{\arcsecond} to \SI{30}{\arcsecond}.}
    \label{fig:spec_g75_n1}
\end{figure}

\begin{figure}[htb!]
    \centering
    \includegraphics[width=0.4\linewidth]{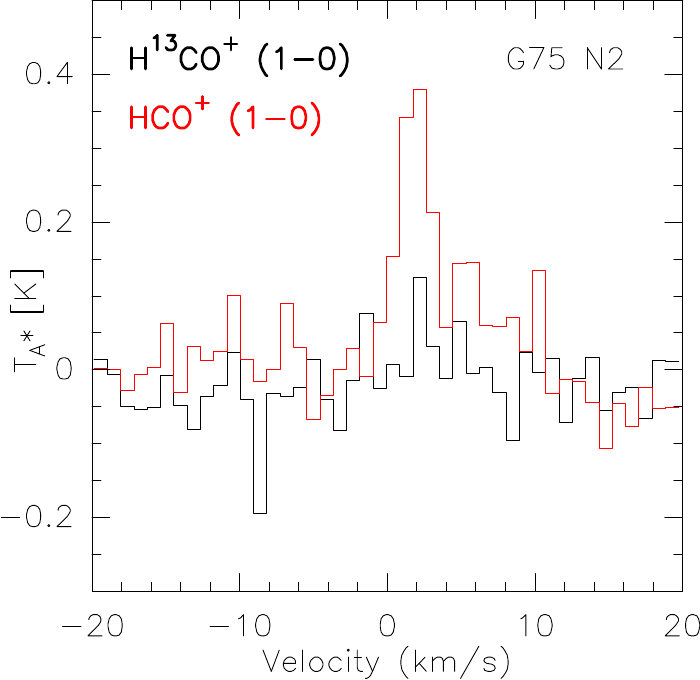}
    \caption{Molecular emission in G75 N2.}
    \label{fig:spec_g75_n2}
\end{figure}

\begin{figure}[htb!]
    \centering
    \includegraphics[width=0.4\linewidth]{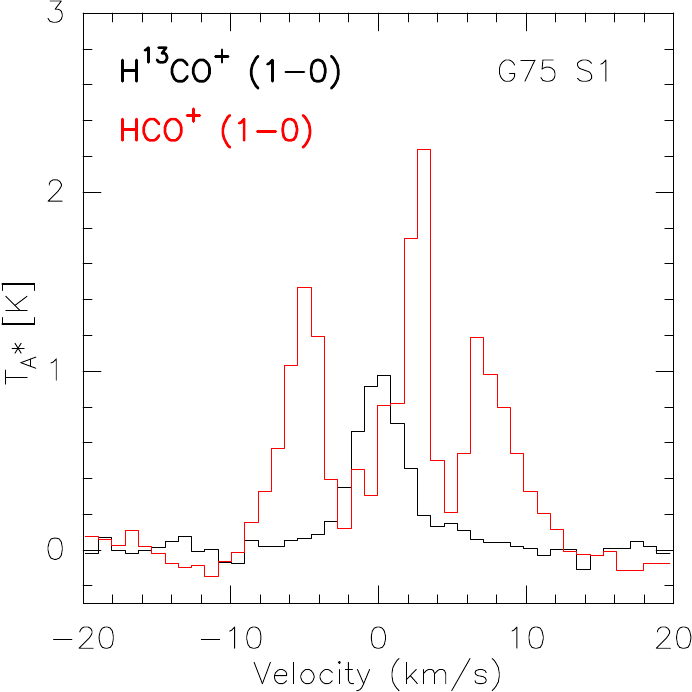}
    \caption{Molecular emission in G75 S1.}
    \label{fig:spec_g75_s1}
\end{figure}

\begin{figure}[htb!]
    \centering
    \includegraphics[width=0.4\linewidth]{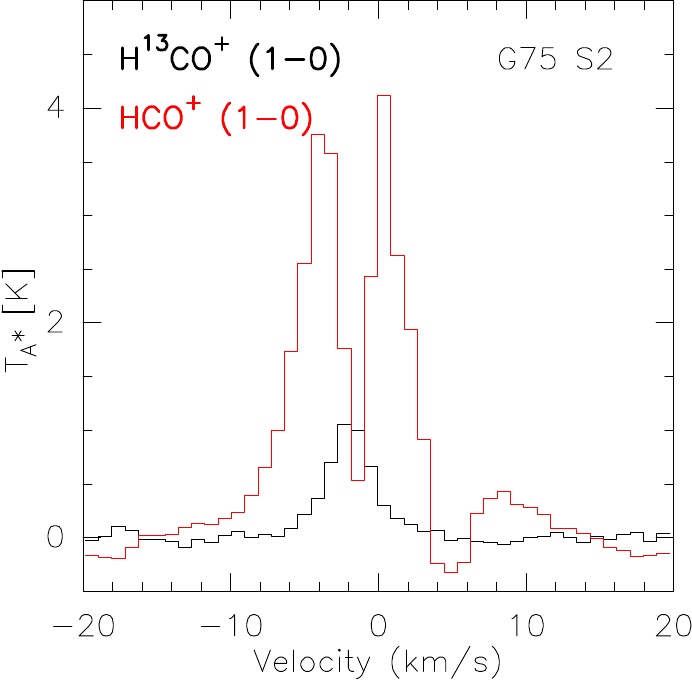}
    \caption{Molecular emission in G75 S2.}
    \label{fig:spec_g75_s2}
\end{figure}

\begin{figure}[htb!]
    \centering
    \includegraphics[width=0.4\linewidth]{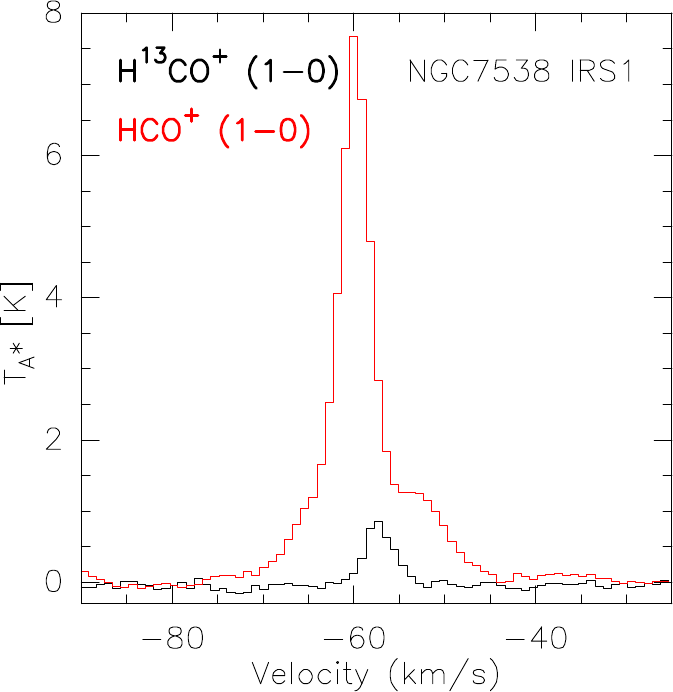}
    \caption{Molecular emission in NGC7538 IRS1.}
    \label{fig:spec_ngc7538_irs1}
\end{figure}

\begin{figure}[htb!]
    \centering
    \includegraphics[width=0.4\linewidth]{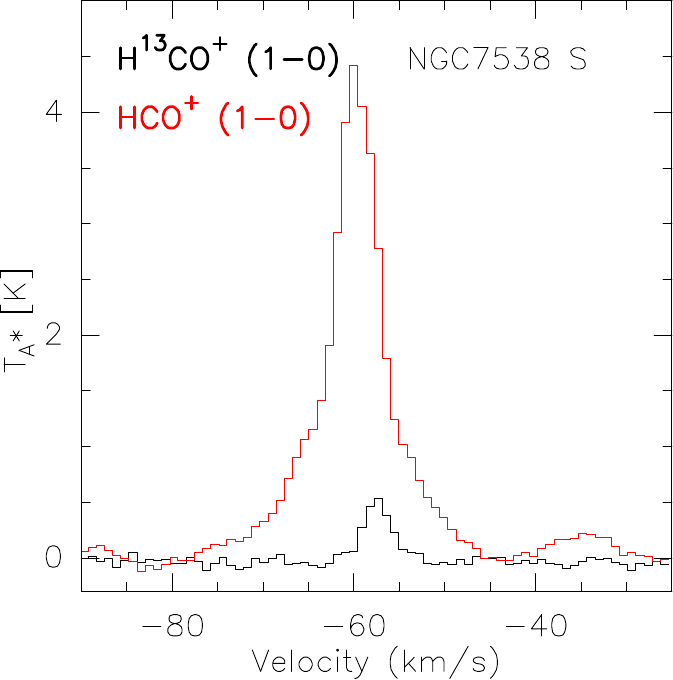}
    \caption{Molecular emission in NGC7538 S.}
    \label{fig:spec_ngc7538_s}
\end{figure}

\begin{figure}[htb!]
    \centering
    \includegraphics[width=0.4\linewidth]{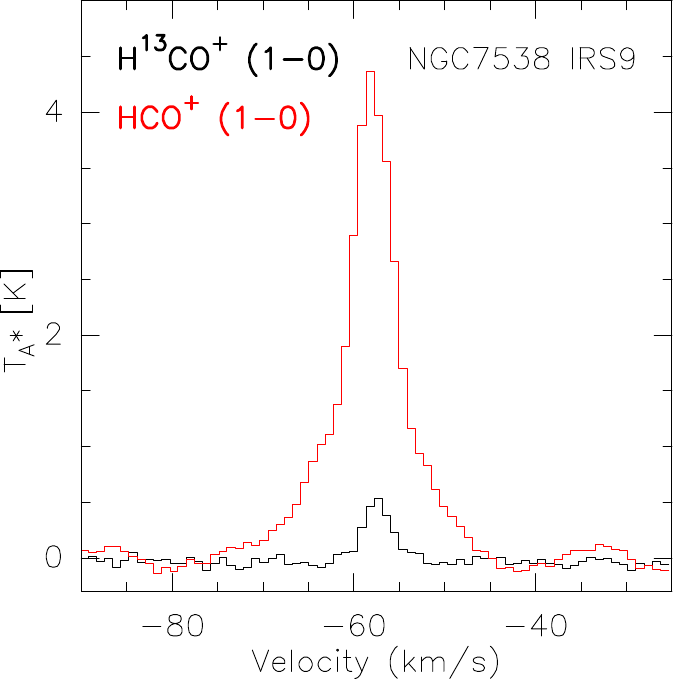}
    \caption{Molecular emission in NGC7538 IRS9.}
    \label{fig:spec_ngc7538_irs9}
\end{figure}

\begin{figure}[htb!]
    \centering
    \includegraphics[width=0.4\linewidth]{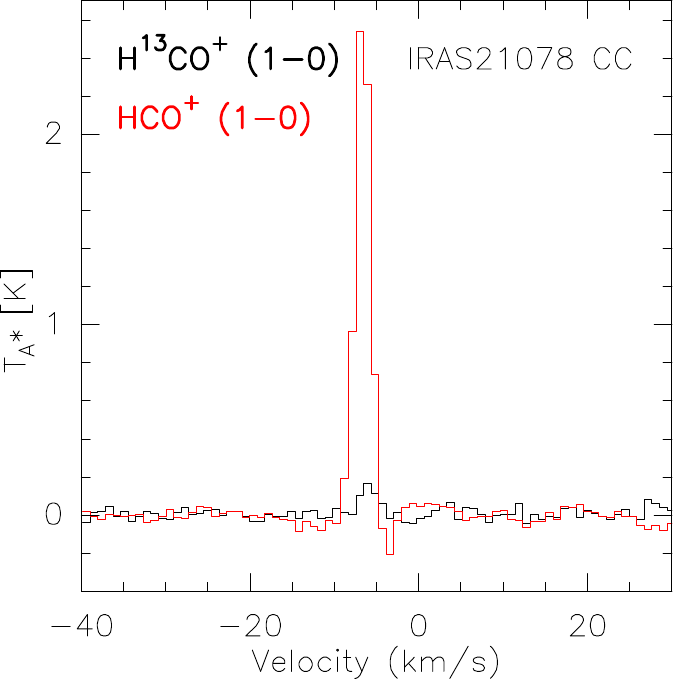}
    \caption{Molecular emission in IRAS21078 CC.}
    \label{fig:spec_iras21078_cc}
\end{figure}

\end{appendix}

\end{document}